\documentclass[12pt,aps,amsmath,latexsym,amsfonts]{JHEP3}
\usepackage{epsfig}
\usepackage{amsfonts,amssymb,amsmath}
\usepackage[hang]{subfigure}
\usepackage[numbers, square, sort&compress]{natbib}

\newcommand{\be}{\begin{equation}}
\newcommand{\ee}{\end{equation}}
\newcommand{\bea}{\begin{eqnarray}}
\newcommand{\eea}{\end{eqnarray}}

\title{Temperature in the Throat}
\author{
Dariush Kaviani\thanks{Email:dariush@ipm.ir}~~and
Amir Esmaeil Mosaffa\thanks{Email:mosaffa@theory.ipm.ac.ir} \\
School of Particles and Accelerators,
Institute for Research in Fundamental Sciences (IPM),
P.O. Box 19395-5531, Tehran, Iran}

\abstract{We study the temperature of extended objects in string theory.
Rotating probe D-branes admit horizons and temperatures a la Unruh effect.
We find that the induced metrics on slow rotating probe D1-branes in holographic
string solutions including warped Calabi-Yau throats have distinct thermal horizons
with characteristic Hawking temperatures even if there is no black hole in the bulk
Calabi-Yau. Taking the UV/IR limits of the solution, we show that the world volume
black hole nucleation depends on the deformation and the warping of the throat. We
find that world volume horizons and temperatures of expected features form not in the
regular confining IR region but in the singular nonconfining UV solution. In the 
conformal limit of the UV, we find horizons and temperatures similar to those on
rotating probes in the AdS throat found in the literature. In this case, we also
find that activating a background gauge field form the $U(1)$ R--symmetry modifies
the induced metric with its temperature describing two different classes of black
hole solutions.}


\preprint{IPM/PA-394}

\begin{document}

\section{Introduction}

Gauge/gravity duality \cite{Maldacena:1997re}
has been at the frontier of research in string theory over the past decade. 
According to gauge/gravity duality, the nearby geometry of a stack of $N$
D3-branes on the smooth ten-dimensional Minkowski space, 
the $AdS_5\otimes S^5$ geometry, is dual to $\mathcal{N}=4$ super Yang-Mills
theory (SYM). This duality implies that $\mathcal{N}=4$ SYM is described at
high temperatures by black holes in $AdS_5$  \cite{Gubser:1996de}. The duality
also led to the construction of more general and realistic holographic string
solutions where some supersymmetry and/or conformal invariance are broken
\cite{Klebanov:2000hb,Klebanov:2000nc,Klebanov:1998hh}
(see also \cite{Herzog:2001xk}). The prototypical example of such holographic
string solutions is the warped throat geometry known as the Klebanov-Strassler
(KS) throat, \cite{Klebanov:2000hb}, produced by placing $N$ regular D3-branes
and $M$ fractional D3-branes (wrapped D5-branes) not at a smooth point but at
a generic Calabi-Yau singularity, the conifold point. In this set up, the
presence of regular and fractional background branes, respectively, reduce the
supersymmetry to $\mathcal{N}=1$ and break conformal invariance, and the
singularity is removed by conifold deformation. Finally, these branes dissolve
into R--R and NS--NS fluxes producing a fully regular supergavity solution with
a large hierarchy, consisting of a warped throat region that smoothly closes off
in the IR, where warping becomes constant. The dual gauge theory is 
$SU(N+M)\times SU(N)$ and includes confinement and chiral symmetry breaking.
Although the exact KS solution is noncompact, it is well approximated by a compact
solution, \cite{Giddings:2001yu}, where the UV end of the throat is attached to the
compact internal Calabi-Yau space. This sets the UV/IR  hierarchy of the throat. Far
from the tip of the throat, the KS solution is well-approximated by the 
Klebanov-Tseytlin (KT) solution, \cite{Klebanov:2000nc}, where warping is continuous
and varies logarithmically. The dual $SU(N+M)\times SU(N)$ gauge theory is chiral, 
but nonconformal and nonconfining including RG cascade. In the limit where the $M$
fractional branes are removed, the KT solution reduces to the Klebanov-Witten (KW)
throat solution,  \cite{Klebanov:1998hh}, dual to $\mathcal{N}=1$ $SU(N)$ 
superconformal gauge field theory. The metrics of these Calabi-Yau throats asymptote
AdS space in the UV and are equivalent in the mid throat region. In the IR region,
however,  the KT and KW solutions are singular at the tip while the KS caps off 
smoothly. Nonextremal generalizations of these holographic backgrounds have been
considered in \cite{Buchel:2000ch} where it was shown that at sufficiently high
temperature the system develops a horizon with a corresponding Hawking temperature
(see also \cite{Herzog:2001xk}).

The extension of holographic backgrounds has also been considered by adding and
using probes, \cite{Karch:2001cw}, to model flavor physics \cite{Karch:2002sh,Sakai:2003wu},
and quantum critical phenomena, \cite{Karch:2007pd}, complementary to other works on charged
$AdS$ black holes, \cite{Herzog:2009xv}. The remarkable property of the probe setup is that
the dynamics of probe branes describe the strongly coupled dynamics of a known field theory.
Furthermore, it has the advantage that the original supergravity background remains unchanged
and the full dynamics is given by the world volume dynamics of the brane which solves easier
than the Einstein equations. The other interesting phenomenon that has been studied in such
setups is the appearance of horizons and temperatures on the world volume of 
{\it{acceleraing probe}} D-branes in the background spacetime 
\citep{Russo:2008gb,Chernicoff:2008sa,Paredes:2008cr,Athanasiou:2010pv,Caceres:2010rm,
Hirata:2008ka,Hirayama:2010xi,Das:2010yw}. The appearance of horizons and temperatures on
accelerated D-branes has been analyzed in \cite{Russo:2008gb} in general terms. The observation
has been that the appearance of horizons and temperatures on accelerating probe branes is in fact 
the usual {\it{Unruh\ effect}}, \cite{Unruh:1976db}, not for point particles but for extended
objects. The presence of horizons on the world-sheet and on time-dependent D-branes in flat
spacetime has been analyzed in \cite{Chernicoff:2008sa,Paredes:2008cr,Athanasiou:2010pv,Caceres:2010rm}
and  \cite{Hirata:2008ka}, respectively. Horizons and temperatures on D-branes from accelerated 
observers in pure AdS spacetime have been studied in \cite{Hirayama:2010xi}. Moreover, in
\cite{Das:2010yw} the appearance of horizons and temperatures has been studied from the world volume
dynamics of probe D-branes embedded in holographic backgrounds of the form $AdS_m\times S^n$, dual to
defect or flavor conformal field theories. The main result of \cite{Das:2010yw} has been that the
induced world volume metrics on rotating probe $D$-branes in the $AdS_5\otimes S^5$ solution have
thermal horizons with characteristic Hawking temperatures even if there is no black hole in the bulk
$AdS$. In the simplest example of a rotating probe D1-brane, the thermal horizon and Hawking temperature
on the world volume of the rotating probe have been found to take a very simple form, proportional to 
the angular velocity $\omega$, given by $r_H=\omega$ and $T_H=\frac{\omega}{2\pi}$. It has been argued
there that these classical solutions describe thermal objects in the dual CFT, including a monopole or
a quark at finite temperature $T_H$. It has also been shown that the application of gauge/gravity duality
to such systems reveals thermal properties such as Brownian motion and AC conductivity in the dual field
theories.

The aim of this work is to extend such previous analyses and study in detail the world volume horizons
and temperatures of rotating probe branes in more general holographic string solutions called warped
Calabi-Yau throats,  \citep{Klebanov:2000hb,Klebanov:2000nc,Klebanov:1998hh}, where some of the
supersymmetry and/or conformal invariance are broken. The motivation is the fact that these holographic
conifold backgrounds have distinct metrics and additional fluxes which modify the IR behavior of the 
$AdS$ spacetime (singular KW, KT, regular KS). The induced world volume metric on the probe in such
backgrounds is thereby expected to admit new examples of world volume black hole geometries characterized
by world volume horizons and temperatures of distinct features. By gauge/gravity duality, these are
expected to describe the temperature in the flavor sector of the corresponding $\mathcal{N}=1$ gauge theory.
In addition, we note that in the UV and in the absence of fractional background branes the solution,
\cite{Klebanov:1998hh}, admits a local $U(1)$ R--symmetry with the associated gauge fields appearing as
fluctuations of the metric and other background fields \citep{Ceresole:1999zs,Kim:1985ez,Gunaydin:1984fk}
(see also \cite{Herzog:2001xk}). Thus, even in the UV where these string solutions asymptote $AdS$, the
additional background gauge fields, when activated, modify the behavior of the background spacetime. 
Therefore the induced world volume metric on the probe is again expected to admit new examples of world
volume black hole geometries described by world volume horizons and temperatures of distinct features.
On the other hand, when the background gauge fields are turned off, one expects to find world volume
horizons and temperatures similar to those of rotating probe branes in the $AdS_5\otimes S^5$ solution,
found in \cite{Das:2010yw}.

The model we consider consists of a slow moving probe D-brane which is rotating freely around spheres
inside warped conifold throats which are at zero temperature. To obtain simple analytic rotating brane
solutions from the action, we consider the simplest probe brane, a D1-brane, and take the small radii
limit as well as the large radii limit of the full supergravity solution, respectively. The former limit
corresponds to the very deep IR region of the KS throat whereas the latter to the UV solutions consisting
of the KT and KW throats. The probe D1-brane, also called D-string or F-string, represents a monopole or
a quark in the dual $\mathcal{N}=1$ gauge theories. We first solve the world volume dynamics of the slow
rotating probe D1-brane in these throat backgrounds and derive the induced world volume metrics. We then
derive from the induced world volume metrics the world volume horizons and temperatures.

We find an assertive affirmative result of these investigations. As with \cite{Das:2010yw} considering the
$AdS_5\otimes S^5$ throat, we find that the induced metrics on the rotating probe in conifold throats also
admit thermal horizons with characteristic Hawking temperatures even if there is no black hole in the bulk.
However, more interestingly we obtain distinct world volume horizons and temperatures and find that world
volume black hole nucleation with horizons and temperatures of expected features in conifold throats depends
on the warping and the deformation of the throat. 

In the very deep IR region of the KS throat where the warping is constant and the solution is regular,
whereby the theory is confining and breaking chiral symmetry, we find that the induced metric on the probe
is not given by the black hole geometry. We find the obstruction to the induced metric for admitting horizons
and temperatures of expected features is due to the confinement or constant warping of the very deep IR. We
derive the radius of the `would be' world volume horizon and obtain a bound on the angular velocities scaling
as the glueball masses. We find the induced radius shrinking continuously, instead of growing, with increasing
the angular velocities. We also vary the parameters and find this behavior robust against such variations.

In the UV solution, including the KT throat where the warping varies logarithmically and the solution is singular,
whereby the theory is nonconfining and chiral, we find, however, distinct world volume horizons and temperatures
of expected features. We find the world volume horizons and temperatures increasing/decreasing continuously with
increasing/decreasing the angular velocities. This behavior is qualitatively no surprise, though we see it here
appearing in the KT and not at the bottom of the KS, as mentioned. However, the special feature we find is that
the world volume horizon is described by the `Lambert transcendental equation', solving to the `Lambert function'.
In our special case of slow rotations, we find that this has a logarithmic series representation, from which we
determine the explicit closed form solution giving the world volume horizon. We find that the world volume horizon
forms about the KT singularity in the IR and appears away in the UV with increasing the angular velocities. The
other, surprising feature we find is that within certain limits in KT the world volume temperatures remain more
or less constant despite varying the world volume horizons. Taking the limits of the world volume temperature,
as the world volume horizon varies inward the UV/IR regions, we find that due to logarithmic warping the world
volume temperature in KT is more or less constant and given in terms of the flux and deformation parameter. 
Another interesting feature we find within the UV/IR limits of KT is a large separation between world volume
temperatures. We find this is due to logarithmic warping by which the flux in the UV can be chosen either large,
or very small, unlike in the IR where the flux has to be large, in order to have a valid SUGRA solution. We also
inspect the parameter dependence of the solution and find the scale and behavior of the world volume horizons and
temperatures in KT subject to certain hierarchies of scales. Taking into account the backreaction of the rotating
solution to the KT SUGRA background, we naturally expect the rotating D1-brane to yield a mini black hole in the
bulk KT. This indicates that the rotating D1-brane describes a thermal object in the dual field theory whereby our
configuration is dual to $\mathcal{N}=1$ gauge theory coupled to a monopole at finite temperature. Since the
$\mathcal{N}=1$ gauge theory itself is at zero temperature while the monopole is at finite temperature we find that
our configuration describes non-equilibrium steady states.

In the UV solution, including the KW throat, $AdS_5\otimes T^{1,1}$, where logarithmic warping is removed, whereby
the theory is conformal, we find that the induced metric on the probe coincides with that of the BTZ black hole with
an angular coordinate suppressed. We find temperatures varying continuously with horizons, similar to those of rotating
probes in $AdS_5\otimes S^5$, \cite{Das:2010yw}, as expected.  Taking into account the backreaction of the rotating
solution to the KW SUGRA background, we naturally expect the rotating D1-brane to yield a mini black hole in the bulk KW.
This shows us that the rotating D1-brane describes a thermal object in the dual conformal field theory whereby our
configuration is dual to $\mathcal{N}=1$ conformal gauge theory coupled to a monopole at finite temperature. Since the
$\mathcal{N}=1$ conformal gauge theory itself is at zero temperature while the monopole is at finite temperature we find
that our configuration describes non-equilibrium steady states. In KW we also consider the case where a massless
background gauge field is activated in $AdS_5$ due to $U(1)$ R--symmetry, including fluctuations in the ten-dimensional
background metric. We choose a nontrivial gauge and solve the supergravity equation of motion describing the gauge field
and obtain  a solution of expected form. We find that our solution consists of a rank one massless field in $AdS$, 
corresponding to a dimension four operator or current in the gauge theory. This is just what would be expected from an
R--current, to which gauge fields correspond. The new interesting feature we find is that in the presence of such a 
background gauge field the incuded world volume metric on the probe has thermal horizons and Hawking temperatures of the
form and behavior similar to those of $AdS$--Reissner--Nordstr\"om and $AdS$--Schwarzschid balck holes in five dimensions.
The special feature we find here is that the Hawking temperature on the probe admits two distinct branches, describing
two classes of black hole solutions. We find that there is one branch which, for large horizon size, the temperature goes
with the quartic of the horizon. The other branch goes at small horizon size as the inverse cube of the cube of the horizon. 
These `small' black holes have the familiar behavior of five-dimensional black holes in asymptotically flat spacetime, as
their temperature decreases with increasing horizon size.

Our paper is organized as follows. In Sec.\,2 we briefly introduce our general basic setup consisting of the general Type
IIB supergravity background and the action of the probe $Dp$-brane in such backgrounds. In  Secs.\,3--5 we consider explicit
examples of type IIB supergravity backgrounds including the KS, KT and KW backgrounds. We first construct rotating probe
D1-brane solutions from the action and derive the induced world volume metrics on the rotating probe D1-brane in these
backgrounds. We then derive and analyze the related world volume horizons and temperatures from the induced world volume
metrics in these supergravity backgrounds. In Sec.\,6 we discuss our results, summarize our findings and conclude with 
future outlook.

\section{General basic set up: IIB theory and Dp-brane action}

For our warped supergavity background, we consider the Calabi-Yau
flux compactification of type IIB theory containing a warped throat
region that smoothly closes off in the IR and is attached to the compact
Calabi-Yau space at the UV end \cite{Giddings:2001yu}.
Type IIB string theory contains NS--NS fields, $\{g_{MN}, B_{2},\Phi \}$,
and R--R forms, $\{C_0, C_2, C_4 \}$. The Type IIB action in the Einstein form
takes the form
\begin{align}
\label{2BEA}
S_{\,\text{IIB}} & =\frac{1}{2\kappa_{10}^2}\int d^{10}x 
\sqrt{|g|}\left(\mathcal{R}_{10}-\frac{\partial_M\tau\cdot \partial^M\bar{\tau}}
{2(\text{Im}\tau)^2}
-\frac{|\,G_{3}|^2}{12\,\text{Im}\tau}-\frac{|\,\tilde{F_{5}}|^2}
{4\cdot5\,!}\right) \notag \\ 
&+\frac{1}{8i\kappa_{10}^2}\int \frac{1}
{\text{Im}(\tau)}C_{4}\wedge G_{3} \wedge G_{3}^*+S_{\,\text{loc}}.
\end{align}
Here $S_{\,\text{loc}}$ stands for localized contributions from 
D-brane and orientifold planes; $G_{3}=F_{3}-\tau H_{3}$
is the combination of R--R and NS--NS field strengths, given respectively as
$F_{3}=dC_{2}$ and $H_{3}=dB_{2}$;
$\tau=C_{0}+ie^{-i\Phi}$ is the axion-dilaton;
$\kappa_{10}^2=\frac{1}{2}(2\pi)^7 {{\alpha}^{\prime}}^4 g_{s}^2$
is the ten-dimensional gravitational coupling with $g_s$ the string coupling
and $\alpha^{\prime}=l_s^2$ the string scale;
$\tilde{F}_{5}=\star_{10}\tilde{F}_{5}=dC_{4}-\frac{1}{2}C_{2}\wedge H_{3}
+\frac{1}{2}B_{2}\wedge F_{3}$ is the self-dual five form with $\star_{10}$ being
the ten-dimensional Hodge-star operator; $\mathcal{R}_{10}$ is the ten-dimensional
Ricci-scalar. The ten-dimensional warped metric and the self-dual five-form read
\cite{Giddings:2001yu}

\begin{eqnarray}
\label{Gmet.}
ds_{10}^2&=& g_{MN}dX^M dX^{N}=h^{-1/2}(y)g_{\mu\nu}dx^{\mu}dx^{\nu}+h^{1/2}(y)
g_{mn}dy^{m}dy^{n},\\ \tilde{F}_{5}&=&(1+\star_{10})\Big[d\alpha(y)\wedge dx^0\wedge dx^1
\wedge dx^2 \wedge dx^3\Big].
\end{eqnarray}
Here $\alpha(y)$ is a scalar quantity and $h(y)$ is the warp factor
with $y^m$ being the internal coordinates. 
The Laplace equation and Bianchi identity imply that the localized
sources to saturate BPS like identity and, $\alpha=h^{-1}$ and
$\star_6 G_3=iG_3$, which implies constant dilaton $\Phi=0$.
Such a string solution is called imaginary self-dual (ISD).

The action of a $Dp$-brane is the sum of the DBI and CS actions,
which in the above background takes the form
\begin{eqnarray}
 \label{DPACTION} S_{Dp}&=&-g_sT_p
\int{d^{p+1}\xi\,e^{-\Phi}\sqrt{-\det(\gamma_{ab}+\mathcal{F}_{ab})}}+g_s
T_p\int{\sum_{p}C_{p+1}\wedge e^{\mathcal{F}_{ab}}}.
\end{eqnarray}
Here $C_{p+1}$ are the IIB R--R background fields coupling to the $Dp$-brane
world volume; $\mathcal{F}_{ab}=\mathcal{B}_{ab}+2\pi\alpha^{\prime}F_{ab}$ is
the gauge invariant field strength with $F_2$ the world volume gauge field and
$\mathcal{B}_2$ the pullback of the NS--NS two-form background field, $B_2$, 
onto the world volume, $\mathcal{B}_{ab}=B_{MN}\partial_a X^M\,\partial_b X^N$, 
$\gamma_{ab}=g_{MN}\partial_a X^M\,\partial_b X^N$ the pullback of the 
ten-dimensional metric $g_{MN}$ in string frame, and $\Phi=0$. Lastly, $\xi^a$
denote the world volume coordinates and $T_p=[(2\pi)^p\, g_s (\alpha^{\prime})^{(p+1)/2}]^{-1}$
is the $Dp$-brane tension.

In what follows we first consider explicit examples of Calabi-Yau throats with
known metric (\ref{Gmet.}) and background fields and determine the explicit form
of the probe $Dp$-brane action, (\ref{DPACTION}), considering a rotating embedding
ansatz for the world volume and assuming that there are no gauge 
fields on the world volume of the probe brane, $F_{ab} = 0$. We then solve the
world volume dynamics from the action, considering slow rotations, and compute the
induced metrics on the probe brane, from which we analyze and derive the world volume
horizons and temperatures.

\section{Temperature in the Klebanov-Strassler throat}

\subsection{The Klebanov-Strassler solution}

As our first example, we take the Klebanov-Strassler (KS) throat geometry
\cite{Klebanov:2000hb} (see also \cite{Herzog:2001xk}), also known as the
warped deformed conifold. The deformed conifold is a nonsingular and  
noncompact Calabi-Yau threefold defined by a hypersurface in $\mathbb{C}^4$ 

\begin{eqnarray}
\label{dccons.}
\sum_{a=1}^4 z_a^2&=&\epsilon^2,\;\;\ z_a\in\mathbb{C}^4,
\end{eqnarray}
and by a radial coordinate

\begin{eqnarray}
\label{dcr.}
\sum_{a=1}^4 |z_a|^2&=&\epsilon^2\cosh\eta=r^3.
\end{eqnarray}
Here $\eta$ is the `radial' coordinate on the conifold and $\epsilon$ is the
deformation parameter of the conifold, which can be made real by phase rotation.
In the limit $\epsilon\rightarrow 0$, Eq.\,(\ref{dccons.}) gives the singular
conifold and describes a cone over a five-dimensional Einstein, base manifold
$T^{1,1}$ of topology $S^{2}\times S^{3}$ with both $S^2$ and $S^3$ spheres
shrinking to zero size at the tip of the cone, $r=0$. The topology of the base
$T^{1,1}$ is parametrized in a standard way by a set of five Euler angles
$\{\theta_i, \phi_i, \psi\}$, where $0\leq \theta_{i}
\leq \pi$, $0\leq \phi_{i} \leq 2\pi$, $0\leq \psi \leq 4\pi$ $(i=1,2)$.
The $S^{2}\times S^{3}$ topology can be then identified as:

\begin{equation}
\label{S2S3}
S^2:\;\;\ \psi=0,\;\;\ \theta_1=\theta_2,\;\;\ \phi_1=-\phi_2\,;\;\;\;\;\;\;\
\text{and} \;\;\;\;\;\;\  S^3: \;\;\ \theta_2=\phi_2=0.
\end{equation}
In the nonsingular limit, at the tip, $\eta\simeq 0$, the $S^2$ shrinks to zero
size while the $S^3$ remains of finite size with radius $\epsilon^{2/3}$ amounting
to the deformation by $\epsilon$, which removes the singularity of the tip. The
deformed conifold contains two independent three-cycles: the $S^3$ at the tip,
known as the  A-cycle, and the Poincare dual three-cycle, known as the B-cycle.

The K\"ahler potential on the deformed conifold derived from (\ref{dcr.})
reads \cite{Candelas:1989js}

\begin{equation}
k(\eta)=\frac{\epsilon^{4/3}}{2^{1/3}}\int_{0}^{\eta}{d\eta^{\prime}
[\sinh(2\eta^{\prime})-2\eta^{\prime}]^{1/3}}
\end{equation}
The metric on the deformed conifold, which is derived from
this K\"ahler potential, takes the form \cite{Candelas:1989js,Minasian:1999tt}

\begin{eqnarray}
\label{dcmetric}
ds_6^2=\frac{1}{2}\epsilon^{4/3}K(\eta)\bigg[ \frac{1}{3K(\eta)^3}
\{d\eta^2+(g^{5})^2\}+\cosh^2\frac{\eta}{2}\{(g^{3})^2+(g^{4})^2\}
\notag \\+ \sinh^2\frac{\eta}{2}\{(g^{1})^2+(g^{2})^2\}\bigg],
\end{eqnarray}
where

\begin{equation}
\label{edef}
g^{1,3}=\frac{e^{1}\mp e^{3}}{\sqrt{2}},\;\;\;\ g^{2,4}=\frac{e^{2}\mp
e^{4}}{\sqrt{2}},\;\;\;\ g^{5}=e^{5}
\end{equation}
with

\begin{align}                                                                  
\label{Basise}
e^{1} & = -\sin\theta_{1} d\phi_{1},\;\;\;\;\;\;\  e^{2}= d\theta_{1}, \;\;\;\;\;\;\ 
e^{3}= \cos\psi \sin\theta_{2}d\phi_{2}-\sin\psi d\theta_{2}, \notag \\
e^{4} &= \sin\psi\sin\theta_{2}d\phi_{2}+\cos\psi d\theta_{2}, \;\;\;\;\;\;\ 
e^{5}= d\psi+\cos\theta_{1}d\phi_{1}+\cos\theta_{2}d\phi_{2},
\end{align}
and

\begin{equation}
K(\eta)=\frac{(\sinh(2\eta)-2\eta)^{1/3}}{2^{1/3}\sinh\eta}.
\end{equation}
In terms of this, the proper radial coordinate which measures the
actual distance up the throat in the six-dimensional, internal metric
is given by

\begin{equation}
\label{prdisKS}
r(\eta)=\frac{\epsilon^{2/3}}{\sqrt{6}}\int_{0}^{\eta}{\frac{dx}{K(x)}}.
\end{equation}

The warping in this background is produced by placing $N$ D3-branes,
sourcing the self-dual R--R five-form field strength $\tilde{F}_5$,
and $M$ D5-branes wrapping a vanishing two-cycle given by the collapsing
$S^2$ sphere, sourcing R--R three-form field strength $F_3$, on the
conifold geometry. There is also a nontrivial NS--NS three-form field,
$H_3$, and the R--R  zero-form, $C_0$, and the dilaton field, $\Phi$,
vanish on this background. The background three-form fluxes of the KS
solution are quantized

\begin{equation}
\frac{1}{(2\pi)^2\,\alpha^{\prime}}\int_{A}{F_3}=M,\;\;\;\;\ \frac{1}
{(2\pi)^2\,\alpha^{\prime}}\int_{B}{H_3}=-K,
\end{equation}
where $M\gg 1$ and $K\gg 1$ denote integers and the total D3-brane charge
is $N=M K$. Due to the presence of these IIB fluxes, 
a large hierarchy of scales can be stabilized, $\epsilon^{1/3}\sim\exp(-2\pi K/3g_sM)$,
\cite{Giddings:2001yu}, and there is a backreaction on the geometry, producing the
ten-dimensional warped line element \cite{Klebanov:2000hb} (see also \cite{Herzog:2001xk})

\begin{equation}
ds_{10}^2=h^{-1/2}(\eta)ds_4^2+h^{1/2}(\eta)ds_6^2.
\end{equation}
Here $h(\eta)$ is the warp factor, $ds_4^2$ is the usual, four-dimensional Minkowski
spacetime metric, and the internal metric $ds_6^2$, Eq.\,(\ref{dcmetric}), is a strongly
warped and deformed throat, which interpolates between a regular $\mathbb{R}^3 \times S^3$
tip in the IR, to an $\mathbb{R} \times T^{1,1}$ cone in the UV. At small $r$ one has:

\begin{equation}
\label{Ksmall}
r\sim\frac{\epsilon^{2/3}}{3^{1/6}\cdot 2^{5/6}},\;\;\;\;\ K\simeq\left(\frac{2}{3}\right)^{1/3},
\end{equation}
where the internal metric (\ref{dcmetric}) along the proper distance (\ref{prdisKS})
smoothly rounds off with a finite $S^3$ of radius $\epsilon^{2/3}/(12)^{1/6}$. Inspection
of the metric (\ref{dcmetric}) in terms of the proper radial coordinate (\ref{prdisKS}) shows
that at large $r$, or for $\eta\simeq 10-15$, the throat explicitly takes the form of a cone
$\mathbb{R} \times T^{1,1}$. Hence $\epsilon^{2/3}$ gives the radius of the nonsingular $S^3$
at the bottom of the throat, and the scale at which the throat asymptotes the $T^{1,1}$ cone:
It sets the IR scale of the geometry. Like in the deformed conifold, the IR geometry is smooth
and the A-cycle is finite in size, with radius $r_A=\sqrt{g_sM\alpha^{\prime}}$, so the 
supergravity approximation remains valid near the tip provided that $g_sM\gg 1$. The other
background fields are \cite{Klebanov:2000hb} (see also \cite{Herzog:2001xk})

\begin{eqnarray}
\label{B2KS}
B_{2}&=&\frac{g_{s}M\alpha^{\prime}}{2}\big[f(\eta)g^{1}\wedge g^{2}+k(\eta)g^{3}\wedge g^{4}\big],
\\ H_{3}&=&\frac{g_{s}M\alpha^{\prime}}{2}\big[d\eta\wedge(f^{\prime}g^{1}\wedge
g^{2}+k^{\prime}g^{3}\wedge g^{4})+\frac{1}{2}(k-f)g^{5}\wedge(g^{1}\wedge
g^{3}+g^{2}\wedge g^{4})\big],\\ \label{C2KS} C_2&=&\frac{M\alpha^{\prime}}{2}\Big[\frac{\psi}{2}
(g^{1}\wedge g^{2}+g^{3}\wedge g^{4})+(1/2-F)(g^{1}\wedge g^{3}+g^{2}\wedge g^{4})
\notag\\ &&-\cos\theta_1\cos\theta_2\, d\phi_1\wedge d\phi_2\Big],
\\ \label{F3} F_{3} &=&\frac{M\alpha^{\prime}}
{2}\big[g^{5}\wedge g^{3}\wedge g^{4}(1-F)+g^{5}\wedge g^{1}\wedge g^{2}\,
F +F^{\prime} d\eta\wedge(g^{1}\wedge g^{3}+ g^{2}\wedge g^{4})\big],
\;\;\;\;\;\;\;\;\ \\ C_4&=&g_s^{-1}\,h^{-1}\wedge dx^0\wedge dx^1\wedge dx^2\wedge dx^3,\\
 \tilde{F}_{5}&=&\mathcal{F}_{5}+\star\mathcal{F}_{5}=B_{2}
\wedge F_{3}+dC_{4}.
\end{eqnarray}
Here we note that $F_3=dC_2$ and $H_3=dB_2$, and
the explicit form of the functions appearing above is
\cite{Klebanov:2000hb} (see also \cite{Herzog:2001xk}):

\begin{eqnarray}
\label{F} F(\eta)&=&\frac{\sinh\eta-\eta}{2\sinh\eta},\\f(\eta)&=&\frac{\eta\cosh\eta-1}
{2\sinh\eta}(\cosh\eta-1), \\ k(\eta)&=&\frac{\eta\cosh\eta-1}{2\sinh\eta}(\cosh\eta+1),
\\ \label{KSWF}h(\eta)&=&2^{2/3}(g_{s}M\alpha^{\prime})^2\,\epsilon^{-8/3}I(\eta),\\
\label{I(eta)} I(\eta)&=&\int_{\eta}^{\infty}
{dx\frac{x\cosh x-1}{\sinh^2x}(\sinh x\cosh x-x)^{1/3}}. 
\end{eqnarray}
The integral (\ref{I(eta)}) cannot be computed analytically, but one 
can readily find the two important limits of the solution for $\eta\rightarrow 0$
and $\eta\rightarrow \infty$ \cite{Klebanov:2000hb} (\cite{Herzog:2001xk}).
As discussed above, near the bottom of the throat (IR), the internal metric takes the form
of an $S^3$ of finite radius whereas
far from the bottom of the throat (UV), the metric takes the form of the KT solution,
asymptoting AdS spacetime (see sections 4 \& 5). The related limits of (\ref{I(eta)})
are \cite{Klebanov:2000hb} (see also \cite{Herzog:2001xk}):

\begin{eqnarray}
\label{I(eta)IR}
I(\eta\rightarrow 0)&\rightarrow& a_0+\mathcal{O}(\eta^2)\;\ \text{with}\;\
a_0\approx 0.71805,\\ \label{I(eta)UV} I(\eta\rightarrow \infty)&\rightarrow&
3\cdot 2^{-1/3}(\eta-1/4)e^{-4\eta/3}.
\end{eqnarray}

In the next sections, we will consider the limits (\ref{I(eta)IR})--(\ref{I(eta)UV}),
in order to obtain simple analytic rotating brane solutions, which compute the induced
world volume metrics, and world volume horizons and temperatures, in turn.

\subsection{Induced metric and Hawking temperature in the KS throat}

We now solve the world volume dynamics of a rotating probe D1-brane
in the KS throat and derive the induced world volume metric from which
we obtain the world volume horizon and temperature in turn. In order to
obtain simple analytic rotating D1-brane solutions consistent with the
IR limit of the supergravity solution, in this section we will consider
the very small radii limit near the bottom of the throat, corresponding
to the very deep IR limit of the KS solution. In this limit, (\ref{I(eta)})
is given by (\ref{I(eta)IR}), and accordingly, 
the warp factor, (\ref{KSWF}), is constant, and given by  
$h_0=a_0(g_sM\alpha^{\prime})^2\,2^{2/3} \epsilon^{-8/3}$
\footnote{In the next sections, we will consider the very large radii limit
where the the explicit form of the warp factor (\ref{KSWF}) is determined by
(\ref{I(eta)UV}).}. Hence the
ten-dimensional background metric (\ref{dcmetric}) takes the form
\cite{Klebanov:2000hb} (see also \cite{Herzog:2001xk})

\begin{eqnarray}
\label{KSmettip}
ds_{10}^2&\rightarrow&\frac{\epsilon^{4/3}}{(2)^{1/3}a_0^{1/2}(g_sM
\alpha^{\prime})}(dx^2-dt^2)\notag\\ &&+a_0^{1/2}\,6^{-1/3}(g_sM
\alpha^{\prime})\left\{\frac{1}{2}d\eta^2+\frac{1}{2}(g^5)^2+(g^3)^2
+(g^4)+\frac{\eta^2}{4}[(g^1)^2+(g^2)^2]\right\}.\notag\\
\end{eqnarray}
We also note that near the bottom of the throat
the $S^2$ sphere shrinks to zero size while the $S^3$ sphere
remains finite, as discussed. To obtain the explicit form of the
metric near the bottom of the throat, we may therefore consider
an $S^3$ round in (\ref{S2S3}). Since we are interested in rotating
solutions in the background (\ref{KSmettip}), we may also fix 
$\theta_1=\theta=\pi/2$ (which together with (\ref{S2S3}) imply
$g^{2,4}=0$). The full background metric then reads

\begin{eqnarray}
\label{10DIRKS}
ds_{10}^2&\rightarrow&\frac{\epsilon^{4/3}}{(2)^{1/3}a_0^{1/2}
(g_sM\alpha^{\prime})}(dx^2-dt^2)\notag\\ &&+(2^{-1}\,a_0^{1/2}
\,6^{-1/3})(g_sM\alpha^{\prime})\left\{d\eta^2+d\psi^2+B(\eta)
d\phi^2\right\},\notag\\
\end{eqnarray}
where $B(\eta)=1+\eta^2/4$. We also note that in this background, (\ref{10DIRKS}),
one has, according to (\ref{F}),
$F\simeq 1/2$, and so from (\ref{C2KS}) and (\ref{B2KS}) one infers that in
the background (\ref{10DIRKS}) (where $g^{2,4}=0$) $C_2$ and $B_2$ are locally
vanishing. 

To evaluate the action of the probe D1-brane in the background (\ref{10DIRKS}),
we need to specify our embedding ansatz. We consider the probe D1-brane extending
in both $t$- and $\eta$-directions, while localized at $x=0$. Furthermore, take it
spinning in the $\psi$- and/or $\phi$-directions. Thus the D1-brane world volume is
specified by $\phi(\eta,t)$ and/or $\psi(\eta,t)$. To evaluate
(\ref{DPACTION}) in the background (\ref{10DIRKS}) for this embedding ansatz, it suffices
to consider the simplest single field case, setting either $\phi(\eta,t)$ or $\psi(\eta,t)$
constant. Setting $\psi=const.$, gives the action of the rotating D1-brane in the form:

\begin{eqnarray}
\label{DBIACKS}
S_{\text{D1}}&=&-\frac{g_s\,T_{D1}\,
\epsilon^{4/3}}{2(12)^{1/3}}\int{dt\,d\eta\,\sqrt{1+B(\eta)(\phi^{\prime})^2
-\frac{a_0 (g_sM\alpha^{\prime})^2\,B(\eta)}{(24\epsilon^4)^{1/3}}\dot{\phi}^2}}.\label{L}
\end{eqnarray}
Here we note that in the limit of small velocities the higher-order non-canonical
kinetic terms in (\ref{DBIACKS}) can be dropped (after Taylor expansion). The Lagrangian and
equation of motion for the slow rotating D1-brane then take the form:

\begin{equation}
L\equiv1+\frac{1}{2}B(\eta)(\phi^{\prime})^2-\frac{a_0 (g_sM\alpha^{\prime})^2\,B(\eta)}
{2(24\epsilon^4)^{1/3}}\dot{\phi}^2,
\end{equation}

\begin{eqnarray}
\frac{\partial}{\partial\eta}\left[B(\eta)\phi^{\prime}(\eta,t)\right]&=&\frac{a_0 (g_sM
\alpha^{\prime})^2}{(24\epsilon^4)^{1/3}}\frac{\partial}{\partial t}\left[B(\eta)
\dot{\phi}(\eta,t)\right].
\end{eqnarray}

Now, consider the rotating solution of the form

\begin{eqnarray}
\phi(\eta,t)=\omega\,t-2\tan^{-1}\left(\frac{\eta}{2}\right).
\end{eqnarray}
Put these into the background and obtain the induced metric on the D1-brane as

\begin{eqnarray}
\label{indmet1KS}
ds_{ind}^2&=&-\frac{a_0^{1/2}(g_sM\alpha^{\prime})\,\Omega(\eta)}{2\cdot 6^{1/3}}
dt^2+\frac{a_0^{1/2}(g_sM\alpha^{\prime})}{2\cdot 6^{1/3}B(\eta)}\left(1+ B(\eta)
\right)d\eta^2\notag\\ &&+\frac{a_0^{1/2}(g_sM\alpha^{\prime})\omega}{2\cdot 6^{1/3}}dt d\eta,
\;\;\ \text{with}\;\;\
\Omega(\eta)=\frac{(24\,\epsilon^4)^{1/3}}{a_0(g_sM\alpha^{\prime})^2}-B(\eta)\omega^2.
\end{eqnarray}

To eliminate the cross term in this metric, introduce a new coordinate

\begin{equation}
\tau=t-\frac{a_0^{1/2}(g_sM\alpha^{\prime})\omega}{2\cdot 6^{1/3}}\int{\frac{d\eta}{\Omega(\eta)}}.
\end{equation}
The metric (\ref{indmet1KS}) then becomes

\begin{eqnarray}
ds_{ind}^2&=&-\frac{a_0^{1/2}(g_sM\alpha^{\prime})\Omega(\eta)}{2\cdot 6^{1/3}}d\tau^2\notag\\
&&+\frac{a_0^{1/2}(g_sM\alpha^{\prime})}{2\cdot 6^{1/3} B(\eta)\Omega(\eta)}
\left\{\left(1+B(\eta)\right)\left[\frac{a_0^{1/2}(g_sM\alpha^{\prime})\Omega(\eta)}
{2\cdot 6^{1/3}}\right]+B(\eta)\omega^2\right\}d\eta^2.\notag\\
\end{eqnarray}

One can go on and find the location of the horizon in the induced metric by the
usual methods, setting $g^{\eta\eta}=0$. However, as we will show in the following,
some unexpected features appear in this case.  First note that this equation has no
solutions for $\eta$ if both angular momenta are zero. Therefore with no rotations
there will be no world volume horizon as expected. Next consider the case with
$\omega\neq0$.  One can see that in the limit of small  $\omega$, the horizon moves
away from the bottom of the throat. As $\omega$ increases, the  horizon moves towards
the bottom and in the limit of large  $\omega$ it will eventually hit the very bottom
of the throat.

This tells us that the world volume black hole is likely to nucleate away from the
bottom of the throat with an horizon that grows by increasing the angular momentum.
The horizon grows in two directions, one away from the bottom of throat and one 
towards it. The former part is outside the regime of validity of the metric we have
considered in this section and the latter is described by the induced metric above.
The feature just described predicts a specific value for  angular momentum for which
the horizon reaches the bottom of the throat. When this happens, the world volume of
the  brane that is rotating in the IR throat region is completely inside the world
volume black hole. 

All of this can be seen by studying the following equation

\begin{equation}
\label{KShor.}
g^{\eta\eta}(\eta_0)=0\rightarrow \eta_0=\left|\frac{2}{\omega}\right|
\sqrt{\frac{(24\,\epsilon^4)^{1/3}}
{a_0(g_sM\alpha^{\prime})^2}-\omega^2}
\end{equation}

This relation already shows the upper bound on the angular velocities

\begin{equation}  
\label{realcond.}
\omega^2<\frac{(24\,\epsilon^4)^{1/3}}{a_0(g_sM\alpha^{\prime})^2}.
\end{equation}

One should have in mind that, as explained above, the bound on
$\omega$ is due to the maximum radius of horizon in the IR region
before hitting the bottom of throat. It is also interesting to
note that according to our bound (\ref{realcond.}) the angular
velocity scales less than or equal to the mass of glueball
and Kaluza-Klein states, given by 
$m_{\text{glueball}}^2\sim m_{\text{KK}}^2\sim \epsilon^{2/3}
/(g_sM\alpha^{\prime})$ \cite{Herzog:2001xk}.
In the special case where the angular velocities scale as glueball
masses and the inequality is saturated, one has $\eta_0=0$. This
occurs precisely at the IR location where the radial coordinate
(\ref{dcr.}), corresponding to $\eta_0=0$, is given by the radius
of the finite $S^3$, $r_0=\epsilon^{2/3}$. In this special case one
may be inclined to conclude that the world volume temperature is
identically zero, and equals the background temperature. However,
in general and as discussed above, Eq.\,(\ref{KShor.}) does not 
describe the world volume horizon at the bottom of the throat since
by Eq.\,(\ref{KShor.}) $\eta_0$ has no consistent and continuous
limits with varying the angular velocity. Note that if 
$\omega \lesssim m_{\text{glueball}}$ then $\eta_0>0$ and
$r_0\lesssim \epsilon^{2/3}$, which is less than the minimum radius
of the nonvanishing $S^3$ at the bottom of the throat. If
$\omega \ll m_{\text{glueball}}$ then $\eta_0\rightarrow \text{large}$,
giving a very large $r_0$, outside the validity rage of the IR metric.
It is also straightforward to consider the alternative embedding ansatz
for the rotating D1-brane, where the worldvolume dynamics is described by
$\psi$ instead of by $\phi$, and show that the induced metric on the
rotating D1-brane will never describe a world volume black hole. This is
due to the fact that in the IR region of the throat, the $\psi$ angle will
have a constant ($\eta$ independent) warp factor. However,  once the linear
velocity in the $\psi$ direction reaches that of speed of light the metric
degenerates as expected.  These features of the world volume theory are 
likely due to the fact that in the metric (\ref{KSmettip}) for small radii
the warp factor approaches constant, unlike the case for large radii where
the KS is well approximated by KT.

\begin{figure}[!ht]
\begin{center}
\epsfig{file=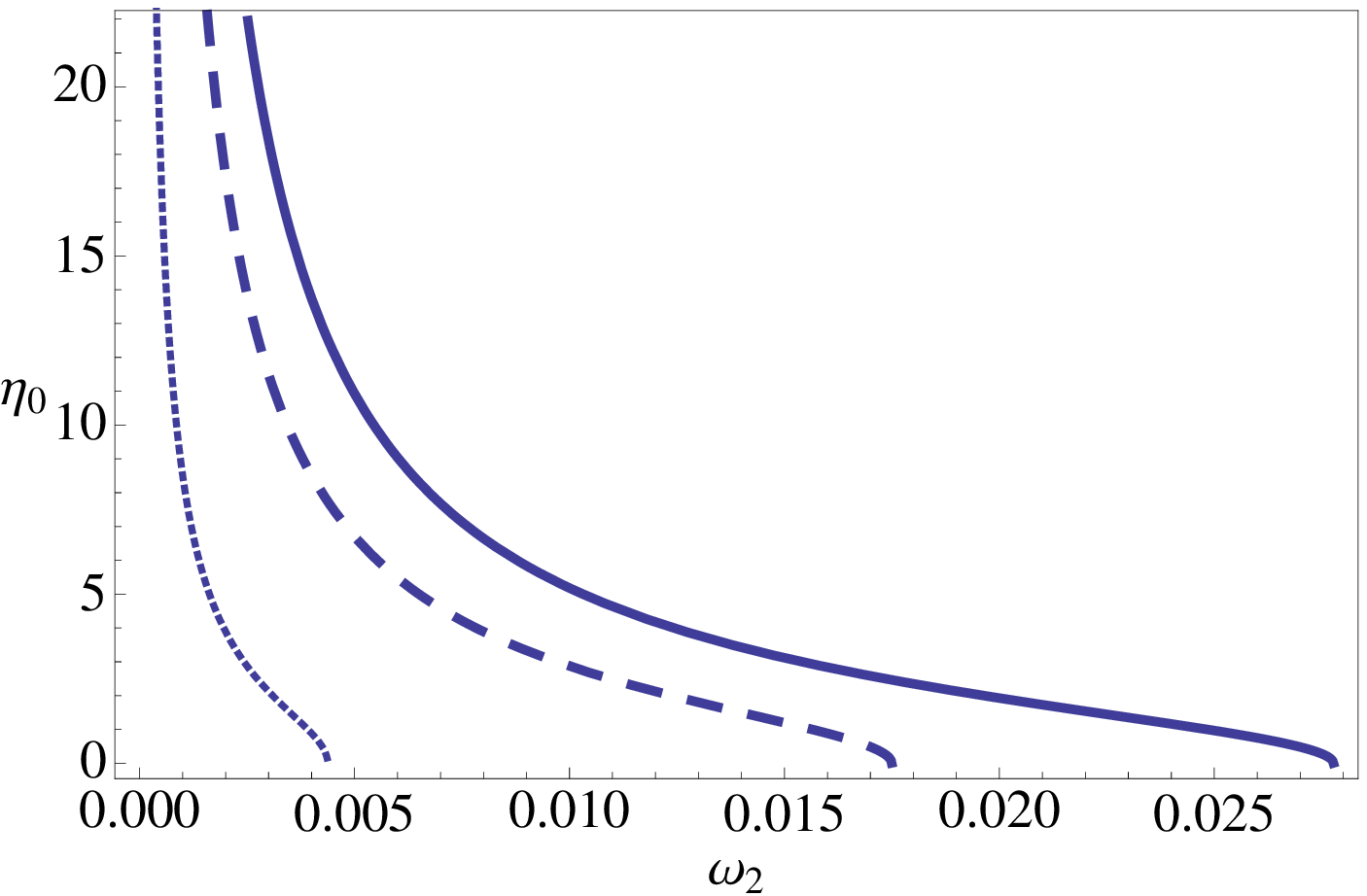,width=.50\textwidth}~~\nobreak
\epsfig{file=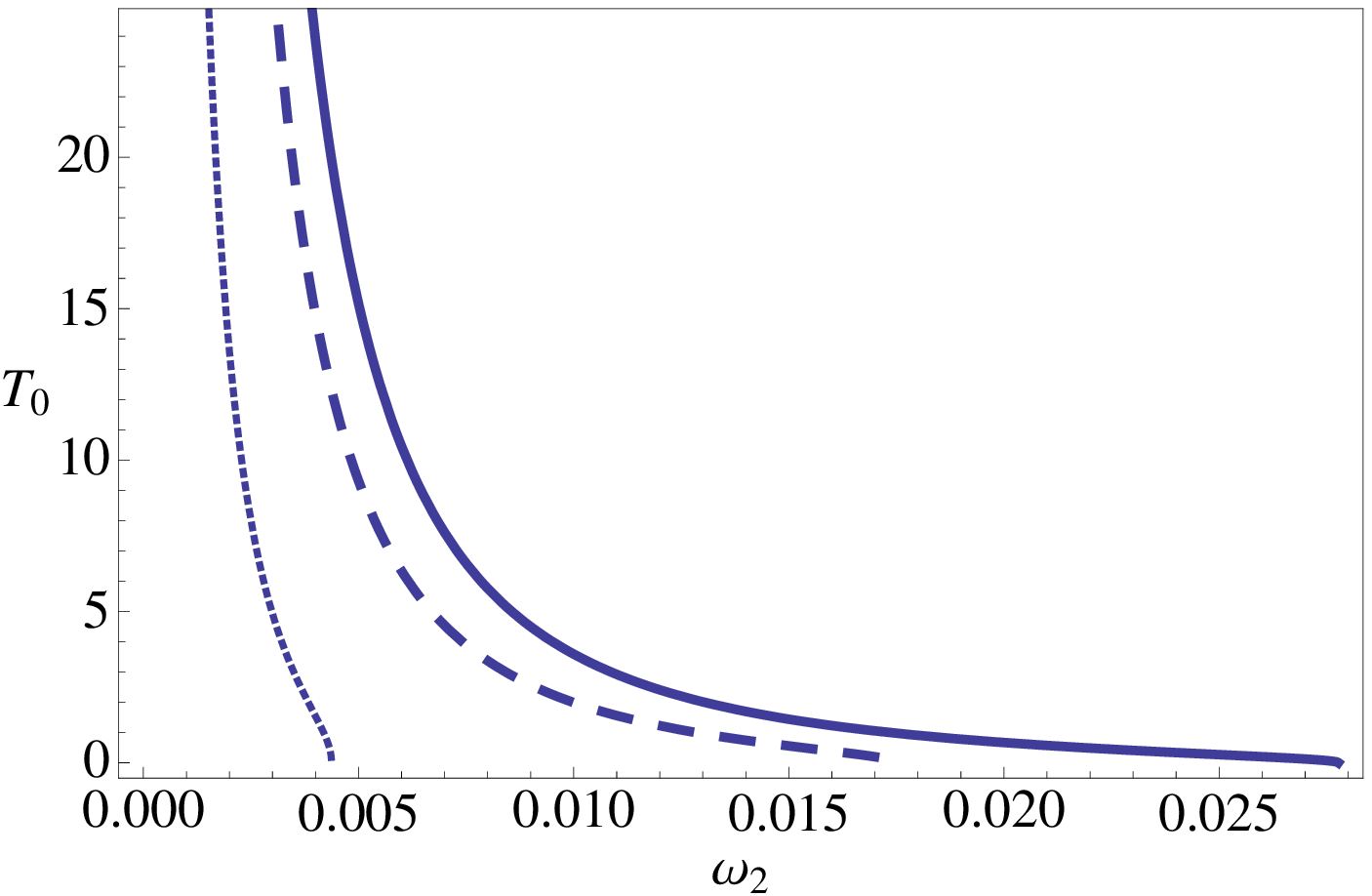,width=.50\textwidth}\\
\epsfig{file=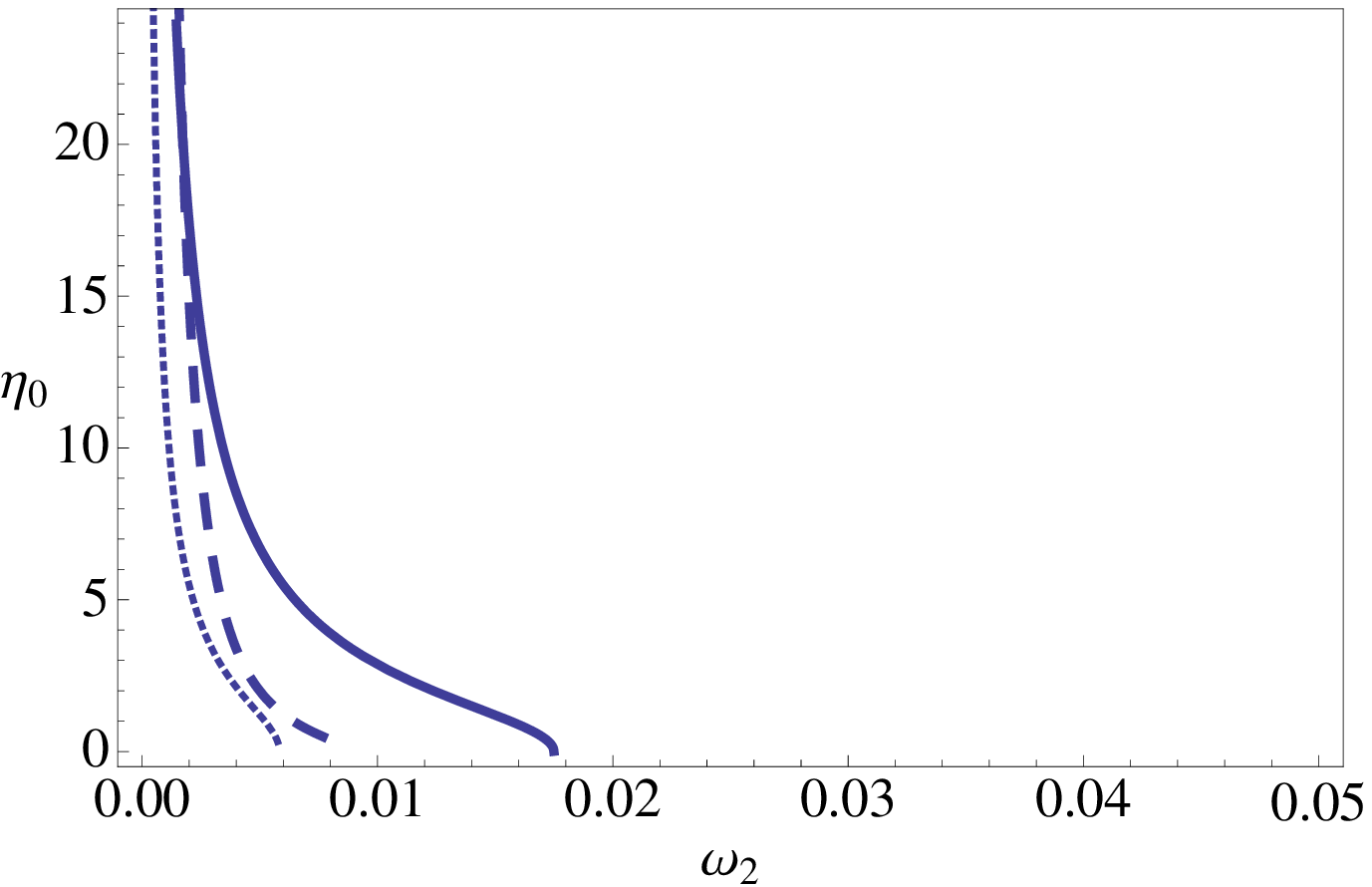,width=.50\textwidth}~~\nobreak
\epsfig{file=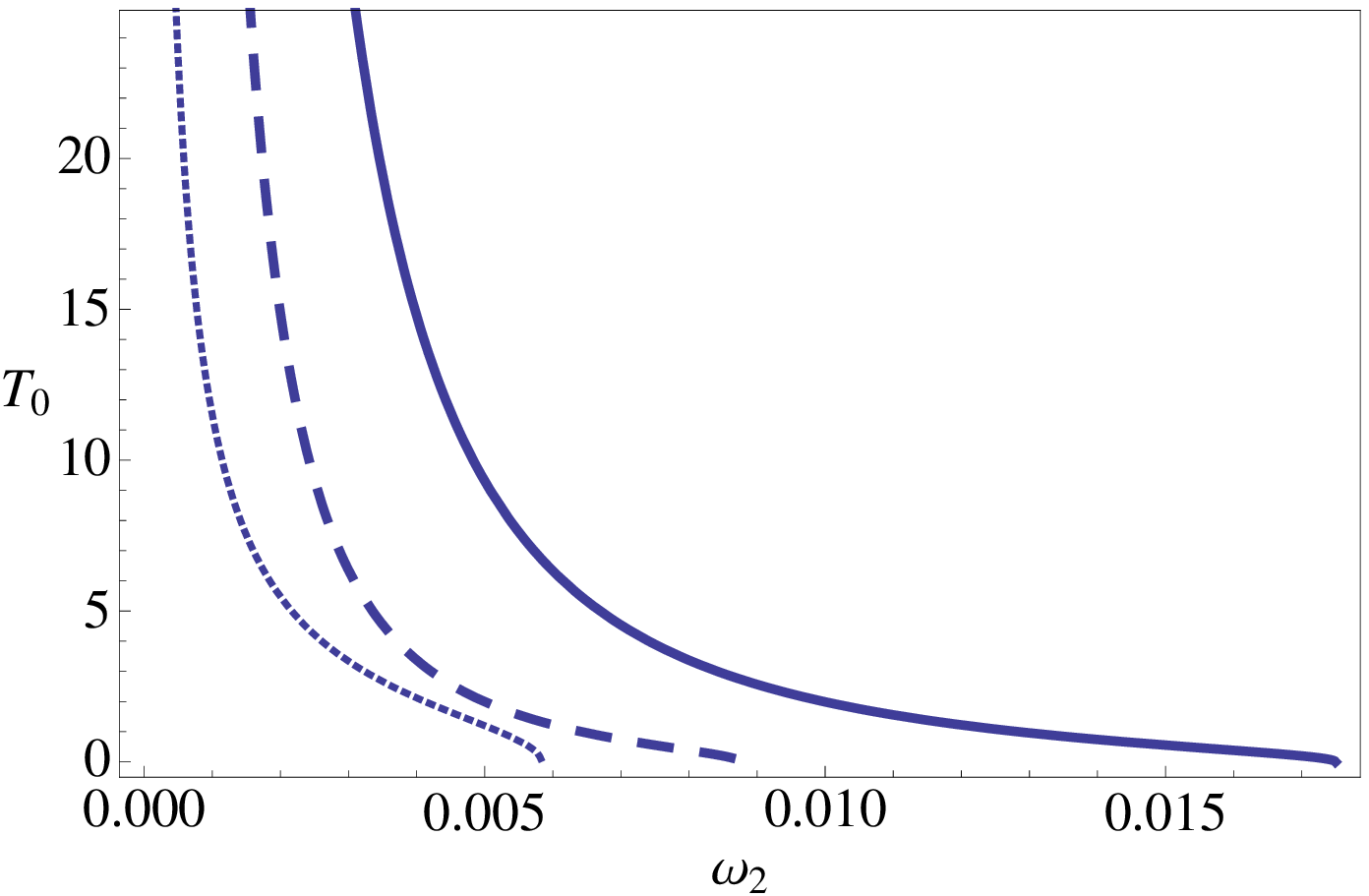,width=.50\textwidth}\\
\caption{[Upper panel] The behaviour of $\eta_0$ and 
$T_0$ for $\epsilon=0.1$ (dotted), $\epsilon=0.8$ (dashed),
$\epsilon=1.6$ (solid). [Lower panel] The behaviour of 
$\eta_0$ and $T_0$ for $\omega_1=10^{-5}$, $\epsilon=0.8$, 
$L^2=3\times 10^2$ (dotted), $L^2=2\times 10^2$ (dashed),
$L^2=10^2$ (solid).}
\label{fig:KSeta0} 
\end{center}
\end{figure}

It is also instructive to inspect the parameter dependence of $\eta_0$ and
$T_0$ by looking at their behavior as a function of $\omega$ for different
values of $\epsilon$ and $g_s M$. Inspection of (\ref{KShor.}) and $T_0$
(see footnote 2) shows that varying $\epsilon$, while keeping $g_sM$ fixed,
only shifts the value of $\omega$ at which $\eta_0$ and $T_0$ hit its minimum
value at the bottom of the throat (see Fig.\,\ref{fig:KSeta0} [upper panel]).
The same behavior appears when doing the opposite; when varying $g_sM$, while
keeping $\epsilon$ fixed (see Fig.\,\ref{fig:KSeta0} [lower panel]). From both
cases it is clear that the behavior of $\eta_0$ and $T_0$ are quite robust
against varying the parameters of the theory, as changing these only changes
the rate of decrease in $\eta_0$ and $T_0$ with $\omega$, but not their
overall, decreasing behavior with increasing $\omega$.

The upshot of this section is that all nontrivial features of the world volume metric
that arise because of accelerating the brane, especially for slow rotations, seem to
occur away from the bottom of the throat which can be explained, for instance, by 
studying the KT throat. This will be the subject of next section\footnote{The nucleation
of black hole away from the bottom of the throat and appearance of an horizon that
approaches the bottom of throat for larger rotations can also be seen by a naive calculation
of the horizon temperature. If one {\it{wrongly}} interprets  $\eta_0$ as appearing in
(\ref{KShor.}) as the radius of horizon, a standard calculation of the horizon temperature
gives 
\begin{equation}
\label{THKS}
T_{\text{0}}=\frac{(g^{\eta\eta})^{\prime}}{4\pi}\bigg|_{\eta=\eta_0}=\frac{6^{1/3}}
{2\pi a_0^{1/2} (g_sM\alpha^{\prime})\,\omega}
\sqrt{\frac{(24\,\epsilon^4)^{1/3}}{a_0(g_sM\alpha^{\prime})^2}-\omega^2}\nonumber.
\end{equation}
This predicts that for larger rotations we will have smaller temperatures
which may seem inconsistent, when compared with the temperature of typical
rotating black hole geometries. However, in other interesting study in the
literature, ref.\,\cite{Nakamura:2013yqa}, using similar D-brane systems,
it has been shown that when an electric field is turned on (in place of 
rotation, or R--charge, considered here), in certain codimensions, the
induced world volume temperature of the probe is given by a decreasing
function of the electric field (see Eq.\,(24) in ref.\,\cite{Nakamura:2013yqa}).}.

\section{Temperature in the Klebanov-Tseytlin throat}

\subsection{The Klebanov-Tseytlin solution}

Far from the tip of the cone, where $\eta$ is large, the deformation of
the conifold can be neglected and Eq.\,(\ref{dccons.}) reduces the
constraint equation of the singular conifold

\begin{equation}
\label{ccons.}
\sum_{a=1}^4 z_a^2=0.
\end{equation}
In this limit, one may introduce another radial coordinate $r$ through

\begin{equation}
\label{r^2KT}
r^2=\frac{3}{2^{5/3}}\epsilon^{4/3}\exp(2\eta/3).
\end{equation}
The K\"ahler potential is then given by \cite{Candelas:1989js}

\begin{equation}
k=\frac{3}{2}\left(\sum_{a=1}^4 |z_a|^2\right)^{2/3}=\frac{3}{2}r^2=\hat{r}^2.
\end{equation}
The metric on the singular conifold, which is derived from this K\"ahler potential,
takes the form \cite{Candelas:1989js}

\begin{equation}
ds_6^2=d\hat{r}^2+\hat{r}^2 ds_{T^{1,1}}\;\;\;\ \text{with} \;\;\;\;\ \label{dsT11}
ds_{T^{1,1}}^2=\frac{1}{9}(g^5)^2+\frac{1}{6}\sum_{i=1}^{4}(g^i)^2,
\end{equation}
where $g^i$'s are given by Eq.\,(\ref{edef})\,-\,(\ref{Basise}). Far
from the bottom of the throat (tip of the cone), where the deformation
parameter can be neglected, by which Eq.\,(\ref{dccons.}) reduces to
Eq.\,(\ref{ccons.}), the KS throat solution joins the Klebanov-Tseytlin
(KT) throat solution \cite{Klebanov:2000nc} 
(see also \cite{Herzog:2001xk}). In this limit, the
ten-dimensional warped line element is that of the KT throat and takes
the form \cite{Klebanov:2000nc} (see also \cite{Herzog:2001xk})

\begin{equation}
\label{KT10Dmet}
ds_{10}^2=h^{-1/2}ds_4^2+h^{1/2}(d\hat{r}^2+\hat{r}^2ds_{T^{1,1}}^2),
\end{equation}
where $ds_{T^{1,1}}^2$ is given by (\ref{dsT11}). The warp factor, $h$, and
other background fields take the form \cite{Klebanov:2000nc}  
(see also \cite{Herzog:2001xk})

\begin{eqnarray}
\label{B2KT}
B_{2}&=&\frac{3g_{s}M\alpha^{\prime}}{4}\bigg[\ln\frac{\hat{r}}{\hat{r}_{\text{UV}}}\bigg]
(g^1\wedge g^2+g^3\wedge g^4), \\ H_{3} &=& dB_2= \frac{3g_{s}M\alpha^{\prime}}{4\hat{r}}d 
\hat{r}\wedge (g^1\wedge g^2+g^3\wedge g^4),\\ \label{C2KT} C_2&\rightarrow&
\frac{M\,\alpha^{\prime}\psi}{2}\,(g^1\wedge g^2+g^3\wedge g^4),
\\ F_{3}&=&dC_2=\frac{M\alpha^{\prime}}{4}g^{5}
\wedge (g^1\wedge g^2+g^3\wedge g^4), \\ \tilde{F}_{5} &=& \mathcal{F}_{5}
+\star\mathcal{F}_{5},\\ \star\mathcal{F}_{5} &=& dC_{4}=g_{s}^{-1}d(h^{-1}(\hat{r}))\wedge dx^{0}
\wedge dx^1 \wedge dx^2 \wedge dx^3, \\ \mathcal{F}_{5} &=& B_{2}\wedge F_{3} = 27
\pi(\alpha^{\prime})^2N_{e}\text{Vol}(T^{1,1}),\\ \label{N_e} 
N_{e}&=& N+\frac{3(g_{s}M)^2}{2\pi}\ln\frac{\hat{r}}{\hat{r}_{\text{UV}}}, 
\\ \label{KTWF} h(\hat{r})&=&
\frac{27\pi{\alpha^{\prime}}^2}{4\hat{r}^4}\left[g_sN+
\frac{3(g_sM)^2}{2\pi}\left(\ln\frac{\hat{r}}{\hat{r}_{\text{UV}}}+\frac{1}{4}\right)\right].
\end{eqnarray}

\subsection{Induced metric and Hawking temperature in the KT throat}

We now solve the world volume dynamics of a slow rotating probe D1-brane
well above the IR limit the KS throat and derive the induced world volume
metric from which we obtain the world volume horizon and temperature in turn.
We obtain simple analytic rotating D1-brane solutions by considering
the very large radii limit, (\ref{I(eta)UV}), away the bottom of the throat,
yet well inside the throat, $\epsilon^{2/3}\ll\hat{r}\ll \hat{r}_{\text{UV}}$,
where the KS throat is well approximated by the UV solution, consisting of
the KT solution. Here we also note that according to (\ref{r^2KT}) for 
$\eta_{\text{UV}}\simeq 10-15$\footnote{See also Section 3, the
discussion below (\ref{Ksmall}).} one has $\hat{r}_{\text{UV}}\simeq
10^2\,\epsilon^{2/3}$, which sets the maximum radial distance for the
throat, the UV scale of the geometry, where the throat is attached to
the compact Calabi-Yau space. In the large-$\eta$ radii limit, (\ref{I(eta)})
is given by (\ref{I(eta)UV}), and, by (\ref{r^2KT}), we may write the warp
factor, (\ref{KTWF}), in the form \cite{Klebanov:2000nc} 
(see also \cite{Herzog:2001xk}):

\begin{equation}
h(\hat{r})=\frac{L^4}{\hat{r}^4}\ln(\hat{r}/\epsilon^{2/3}),\;\;\;\;\
L^4=\frac{81(g_sM\alpha^{\prime})^2}{8},
\end{equation}
giving the ten-dimensional background metric (\ref{KT10Dmet}) of the form
\cite{Klebanov:2000nc} (see also \cite{Herzog:2001xk})

\begin{equation}
ds^2=\frac{\hat{r}^2}{L^2\sqrt{\ln(\hat{r}/\epsilon^{2/3})}}(dx^2-dt^2)+\frac{L^2
\sqrt{\ln(\hat{r}/\epsilon^{2/3})}}{\hat{r}^2}d\hat{r}^2+\frac{L^2}{\hat{r}^2}
\sqrt{\ln(\hat{r}/\epsilon^{2/3})}ds_{T^{1,1}}^2.
\end{equation}
Note that due to logarithmic dependence this metric becomes singular at 
$\hat{r}=\epsilon^{2/3}$. Also, note that because $T^{1,1}$ expands slowly
toward large $\hat{r}$, the curvatures decrease there so that corrections
to the supergravity solution become negligible. Thus even when $g_sM$ is
very small, the supergravity solution considered here is reliable for
sufficiently large radii where $g_sN_{e}\gg 1$ (see Eq.\,(\ref{N_e}))
\cite{Klebanov:2000nc} (see also \cite{Herzog:2001xk})

Considering the same $S^3$ cycle as in the previous section, we obtain
this full background metric of the form

\begin{equation}
\label{10DKTUV}
ds_{10}^2=\frac{\hat{r}^2}{L^2\sqrt{\ln(\hat{r}/\epsilon^{2/3})}}
(dx^2-dt^2)+\frac{L^2\sqrt{\ln(\hat{r}/\epsilon^{2/3})}}{\hat{r}^2}
\left(d\hat{r}^2+\frac{\hat{r}^2}{6}d\phi^2+\frac{\hat{r}^2}{9}d\psi^2\right).
\end{equation}
We also note that in the background (\ref{10DKTUV}), from (\ref{C2KT})
and (\ref{B2KT}) one infers that in the background (\ref{10DKTUV}) 
(where $g^{2,4}=0$) $C_2$ and $B_2$ are locally vanishing.

To evaluate the action of the probe D1-brane (\ref{10DKTUV}), we consider
the same embedding ansatz as before, but now working with the radial
coordinate $\hat{r}$. Thus the D1-brane world volume is again specified by
$\phi(\hat{r},t)$ and/or $\psi(\hat{r},t)$. To evaluate (\ref{DPACTION})
in the background (\ref{10DKTUV}) for the embedding ansatz, it suffices to
set either $\phi$ or $\psi$ constant. Setting $\phi=const.$,
gives the action of the rotating D1-brane in the form:

\begin{eqnarray}
\label{DBIACKT}
S_{\text{D1}}&=&-g_s\,T_{D1}\int{dt\,d\hat{r}\,\sqrt{1+
\frac{\hat{r}^2(\psi^{\prime})^2}{9}-\frac{L^4}{9\,\hat{r}^2}\ln\left(\hat{r}
/\epsilon^{2/3}\right)\dot{\psi}^2}}.
\end{eqnarray}
As before, we note that in the limit of small velocities the higher-order
non-canonical kinetic terms in (\ref{DBIACKT}) can be dropped (after Taylor
expansion). The Lagrangian and equation of motion for the slow rotating D1-brane
then take the form:

\begin{equation}
L=1+\frac{\hat{r}^2(\psi^{\prime})^2}{18}-\frac{L^4}{18\,\hat{r}^2}
\ln\left(\hat{r}/\epsilon^{2/3}\right)\dot{\psi}^2,
\end{equation}

\begin{eqnarray}
\frac{\partial}{\partial\hat{r}}\left[\frac{\hat{r}^2\psi^{\prime}(\hat{r},t)}
{9}\right]&=&\frac{\partial}{\partial t}\left[\frac{L^4}{9\hat{r}^2}\ln\left(
\hat{r}/\epsilon^{2/3}\right)\dot{\psi}(\hat{r},t)\right].
\end{eqnarray}

Consider the simple rotating solution as

\begin{eqnarray}
\psi(\hat{r},t)=\omega\,t-\frac{\omega}
{\hat{r}}+\psi_0.
\end{eqnarray}
Putting this into the background, gives the induced metric on the brane as

\begin{eqnarray}
\label{indmet1}
ds_{ind}^2&=& -\frac{\left[\hat{r}^2-L^4\,\ln(\hat{r}/\epsilon^{2/3})
\overline{\omega}^2\right]}{\sqrt{L^4\ln\left(\hat{r}/\epsilon^{2/3}
\right)}}dt^2\notag+\sqrt{L^4\ln\left(\hat{r}/\epsilon^{2/3}\right)}
\left(\frac{1}{\hat{r}^2}+\frac{\overline{\omega}^2}{\hat{r}^4}\right)
d\hat{r}^2\notag\\ && + \frac{2\overline{\omega}^2}{\hat{r}^2}
\sqrt{L^4\ln\left(\hat{r}/\epsilon^{2/3}\right)}dt\,d\hat{r},
\;\;\;\ \text{with}\;\;\;\ \overline{\omega}^2=\frac{\omega^2}{9}.
\end{eqnarray}
To eliminate the cross term in this metric, we may
consider a coordinate transformation of the form

\begin{equation}
\tau=t-\overline{\omega}^2\int{\frac{d\hat{r}}{\hat{r}^2(\hat{r}^2
-L^4\ln(\hat{r}/\epsilon^{2/3})\overline{\omega}^2)}}.
\end{equation}
The induced metric (\ref{indmet1}) then takes the form

\begin{eqnarray}
\label{indmet2}
ds^2_{ind}&=& -\frac{\left[\hat{r}^2-L^4\,\ln(\hat{r}/\epsilon^{2/3})
\overline{\omega}^2\right]}{\sqrt{L^4\ln\left(\hat{r}/\epsilon^{2/3}
\right)}}d\tau^2+\sqrt{L^4\ln\left(\hat{r}/\epsilon^{2/3}\right)}
\left[\frac{\overline{\omega}^2+\hat{r}^2-L^4\ln(\hat{r}/\epsilon^{2/3})
\,\overline{\omega}^2}{\hat{r}^2(\hat{r}^2-L^4\ln(\hat{r}/\epsilon^{2/3})
\,\overline{\omega}^2)}\right]d\hat{r}^2.\notag\\
\end{eqnarray}
Here we note that, up to $\ln$-dependence, the
metric (\ref{indmet2}) has the form given by the BTZ black hole
with the angular coordinate suppressed (see next sec. Eq.\,(\ref{indmet2KW})).

\begin{figure}[!ht]
\begin{center}
\epsfig{file=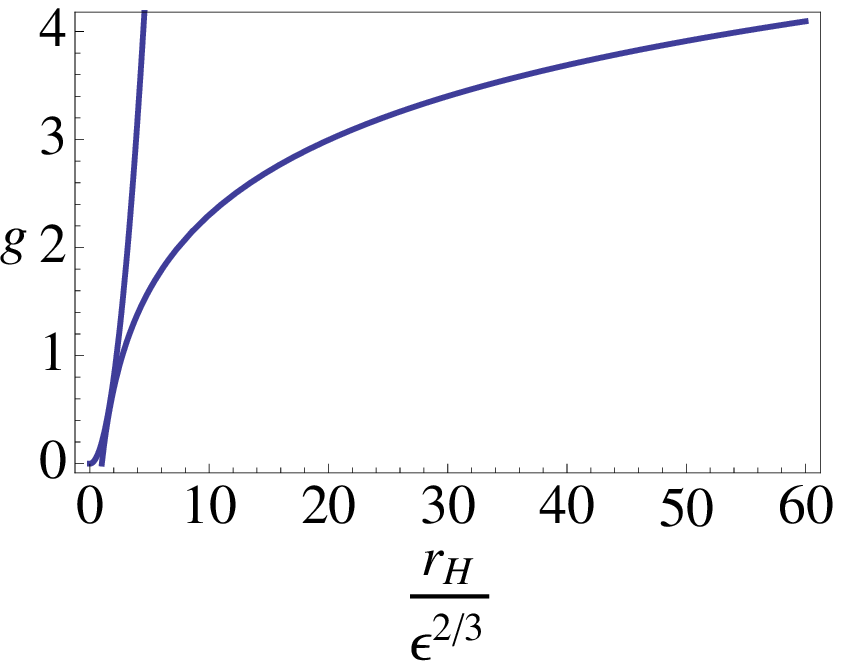,width=.45\textwidth}~~\nobreak
\epsfig{file=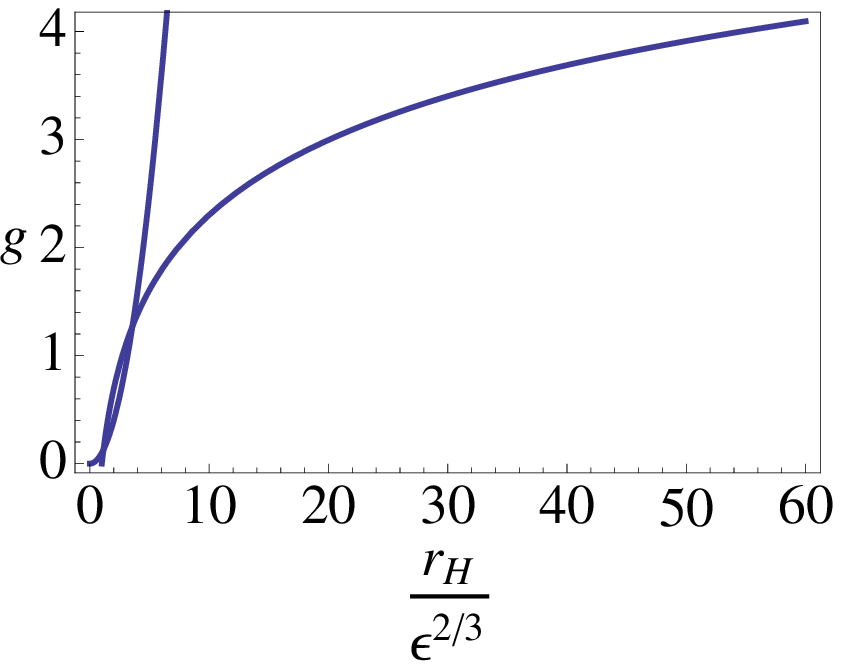,width=.45\textwidth}\\
\epsfig{file=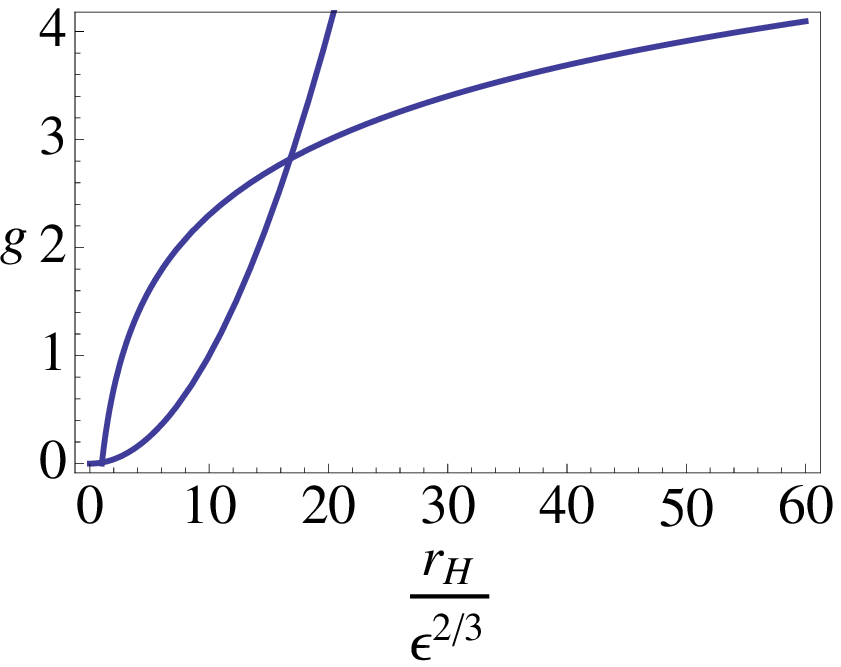,width=.45\textwidth}~~\nobreak
\epsfig{file=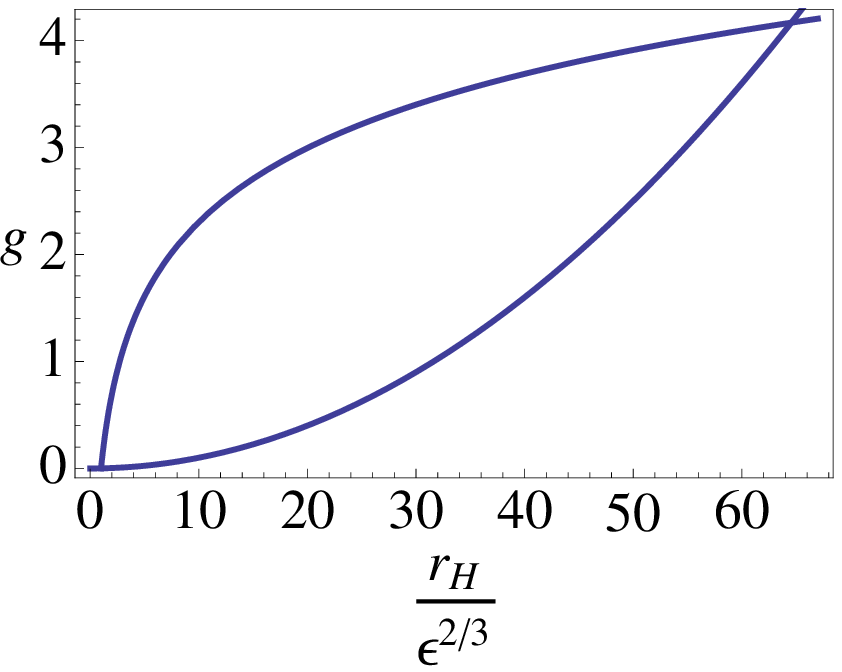,width=.45\textwidth}
\caption{[Upper panel] The plots of $g$ for
$\overline{\omega}^2=\epsilon^{4/3}/L^4$ (left),
$\overline{\omega}^2=10\epsilon^{4/3}/L^4$ (right). 
[Lower panel] The plots of $g$ for $\overline{\omega}^2=50\epsilon^{4/3}/L^4$
(left), $\overline{\omega}^2=100\epsilon^{4/3}/L^4$ (right). The zeros of $g$,
given by the intersecting points, represent the location of the world volume
horizon.} 
\label{fig:KTzs}
\end{center}
\end{figure}

To obtain the horizon, we set from this metric $g^{\hat{r}\hat{r}}=0$, which gives:

\begin{equation}
\label{horkteq.}
g=\hat{r}_H^2-L^4\,\overline{\omega}^2\,\ln(\hat{r}_H/\epsilon^{2/3})=0.
\end{equation}
This equation can have at most two (real positive) zeros. The value
and number of these zeros depends on the value of the conserved charge.
Clearly, at $\overline{\omega}=0$ no zero of (\ref{horkteq.})
appears and there will be no horizon. Now let $\overline{\omega}_c$
be the minimum, critical value of conserved charge for which at least one zero 
of (\ref{horkteq.}) appears. Inspection of (\ref{horkteq.}) shows that for 
$\overline{\omega}_c\simeq\,\epsilon^{2/3}/L^2$, the equation (\ref{horkteq.})
will have one solution in the IR region, close to the tip of the throat
and near the singularity of KT, $\hat{r}_c\simeq\epsilon^{2/3}$.
We can also see this more directly by obtaining an explicit solution of 
Eq.\,(\ref{horkteq.}). By expanding around $\overline{\omega}=0$ and
consider the leading terms, we obtain the solution\footnote{Here we note that
Eq.\,(\ref{horkteq.}) can be rearranged into a Lambert's transcendental equation
of the form $\ln\hat{X}_H=(\epsilon^{4/3}/L^4\,\overline{\omega}^2)\hat{X}_H^2$
whose solution is given in terms of Lambert W-function as $\hat{X}_H=\exp[-W(-2
\epsilon^{4/3}/L^4\,\overline{\omega}^2)/2]$. The expansion of the Lambert
W-function $W(X)$ about $X=\infty$, or $\overline{\omega}=0$, is $W(X)=-\ln[1/X]
-\ln[-\ln[1/X]]-\ln[-\ln[1/X]]/\ln[1/X]^2-\cdots$. Considering the first two
leading terms in this expansion and replacing $X$ by the argument of the exponent
of $\hat{X}_H$ one obtains (\ref{KThor.}), after rearranging back.}:

\begin{equation}
\label{KThor.}
\hat{r}_{\text{H}}=L^2\,\overline{\omega}\sqrt{\left|\ln(\sqrt{2}\,
\epsilon^{2/3}/L^2\,\overline{\omega})\right|}.
\end{equation}
Here again it is clear that for $\overline{\omega}_c\simeq\,\epsilon^{2/3}/L^2$
we get $\hat{r}_c\simeq\epsilon^{2/3}$. Now, increase $\overline{\omega}$.
For any value of conserved charge $\overline{\omega}>\overline{\omega}_c$,
we will have two solutions for (\ref{horkteq.}), one smaller than $\hat{r}_c$ and
one larger, denoted by $\hat{r}_< $ and $\hat{r}_>$ respectively. Increasing
the value of $\overline{\omega}$ continuously, further decreases (increases) the
value of $\hat{r}_< $ ($\hat{r}_>$). One may first be tempted to interpret this 
result as the appearance of a double horizon. However, one should keep in mind that 
the zero at $\hat{r}_< $ is located about the KT singularity away from the UV. 
The conclusion of this analysis is that by rotating the brane inside
the throat, a world volume black hole nucleates around the KT singularity and by increasing
the angular momentum the horizon grows  in size. For rotations $\overline{\omega}\simeq(10^2
-10^3)^{1/2}\,\epsilon^{2/3}/L^2$, the horizon  approaches the UV region, far from the tip
and the KT singularity, and will be of the size of the UV scale of the geometry, 
$\hat{r}_H\rightarrow10^2\epsilon^{2/3}$ (see Fig.\,\ref{fig:KTzs}).

To obtain the Hawking temperature, we Wick-rotate $\tau$ into a Euclidean time, 
and after a straightforward calculation we get:

\begin{eqnarray}
\label{KTtem.}
T_{\text{H}}=\frac{(g^{\hat{r}\hat{r}})^{\prime}}{4\pi}
\bigg|_{\hat{r}=\hat{r}_{\text{H}}}&=&\frac{\hat{r}_{\text{H}}(2\hat{r}_H^2-L^4\,
\overline{\omega}^2)}{4\pi(\overline{\omega}L)^2
\sqrt{\ln\left(\hat{r}_{\text{H}}/\epsilon^{2/3}\right)}},
\end{eqnarray}
where $\hat{r}_H$ is the horizon discussed above. Inspection of (\ref{KTtem.})
shows that the temperature of the black hole solution on the probe, $T_H$, is always
real, positive definite and finite. This is because within intermediate scales 
$\hat{r}_H$ neither hits the KT singularity in the IR nor extends beyond the UV
cutoff, with the temperature (\ref{KTtem.}) scaling roughly as $T_H\gtrsim L^2
\,\epsilon^{2/3}$ in the UV limit where $\hat{r}_H\rightarrow 10^2\epsilon^{2/3}$,
and as  $T_H\lesssim L^2\,\epsilon^{2/3}$ in the IR limit where $\hat{r}_H
\rightarrow\epsilon^{2/3}$.  This means that away from the mid throat region, where
$\hat{r}_H$ approaches the UV/IR ends of the throat, the
temperature of the probe, $T_H$, is uniformly continuous and more or less constant. 
One can also see that $T_H$ increases/decreases continuously with expanding/shrinking $\hat{r}_H$, as
expected. 

If we consider the backreaction of the above solution to the KT supergravity background,
it is natural to expect such D1-branes to form mini black holes in the bulk KT. This
indicates that the rotating D1-brane describes  thermal object with temperature $T_H$ in
the dual field theory. The configuration is dual to $\mathcal{N}=1$ gauge theory coupled
to a monopole at finite temperature. The $\mathcal{N}=1$ gauge theory is itself at zero
temperature while the monopole is at finite temperature $T_H$. Thus such configurations
are in non-equilibrium steady states.

The acceptable values of $T_H$ follow from the validity range for $\hat{r}_H$.
In addition, since the KT
supergravity solution is reliable for sufficiently large radii even when $g_sM\ll 1$,
the temperature scales smaller, or larger, depending on whether $g_sM\gg 1$ (making
$L^2\gg 1$), or $g_sM\ll 1$ (making $L^2\ll 1$) is considered.  In particular, since
the KT solution is singular and consistent with very small deformations
($\epsilon\ll 1$), if $g_sM\ll 1$ then $T_H\gtrsim L^2 \epsilon^{2/3}$ is vanishingly small.
Conversely, if $g_sM\gg 1$ then $T_H$ is finitely (nonvanishingly) small, when $g_sM$ is large
enough. It is conclusive then that the size of $T_H$ in KT at sufficiently large radii, in the
far UV limit, depends on the choice of flux. On the other hand, in the IR limit, we may only 
consider $g_sM\gg 1$, in which case $T_H\lesssim L^2\,\epsilon^{2/3}$ is always finitely small
for large enough $g_sM$, similar to $T_H$ in the far UV limit, when $g_sM$ is large enough. 
Hence it is also conclusive that in KT, when $g_sM\gg 1$, the size of $T_H$ away from the mid
throat region, in the UV/IR limits of $\hat{r}_H$, is essentially same, always finitely small.

\begin{figure}[!ht]
\begin{center}
\epsfig{file=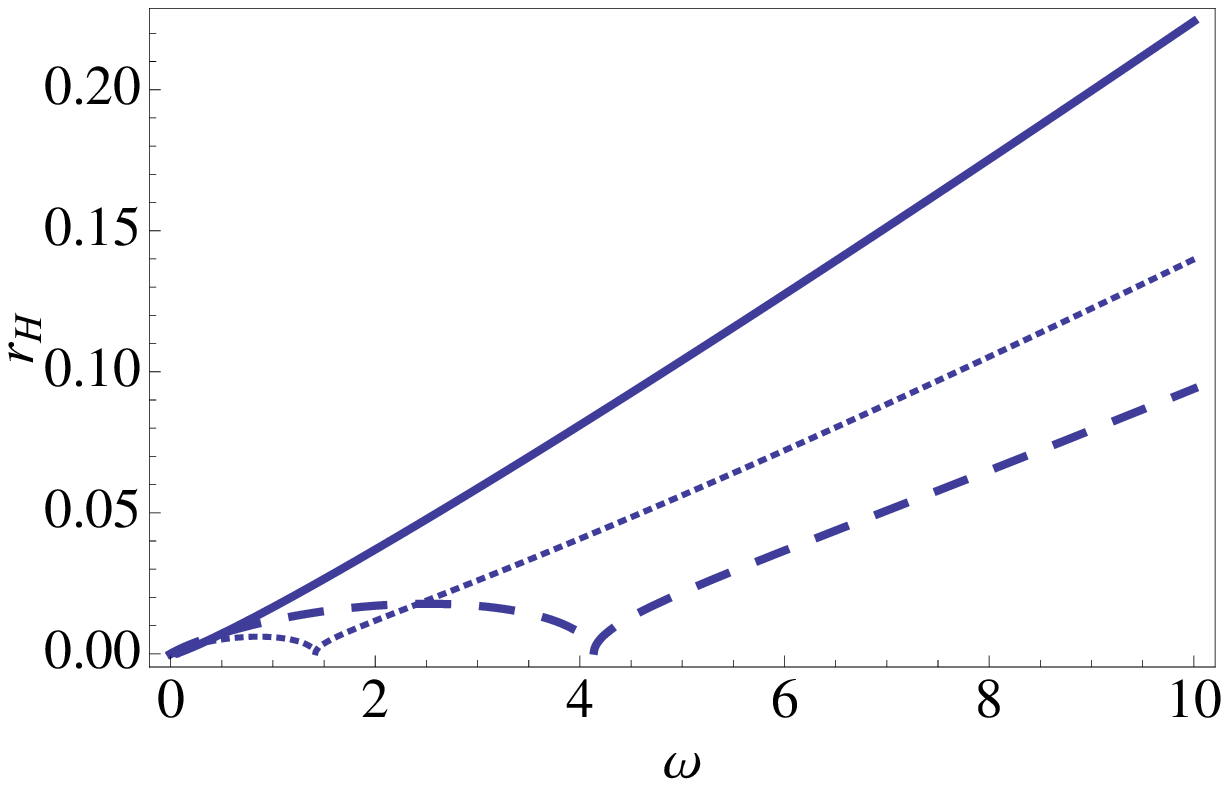,width=.50\textwidth}~~\nobreak
\epsfig{file=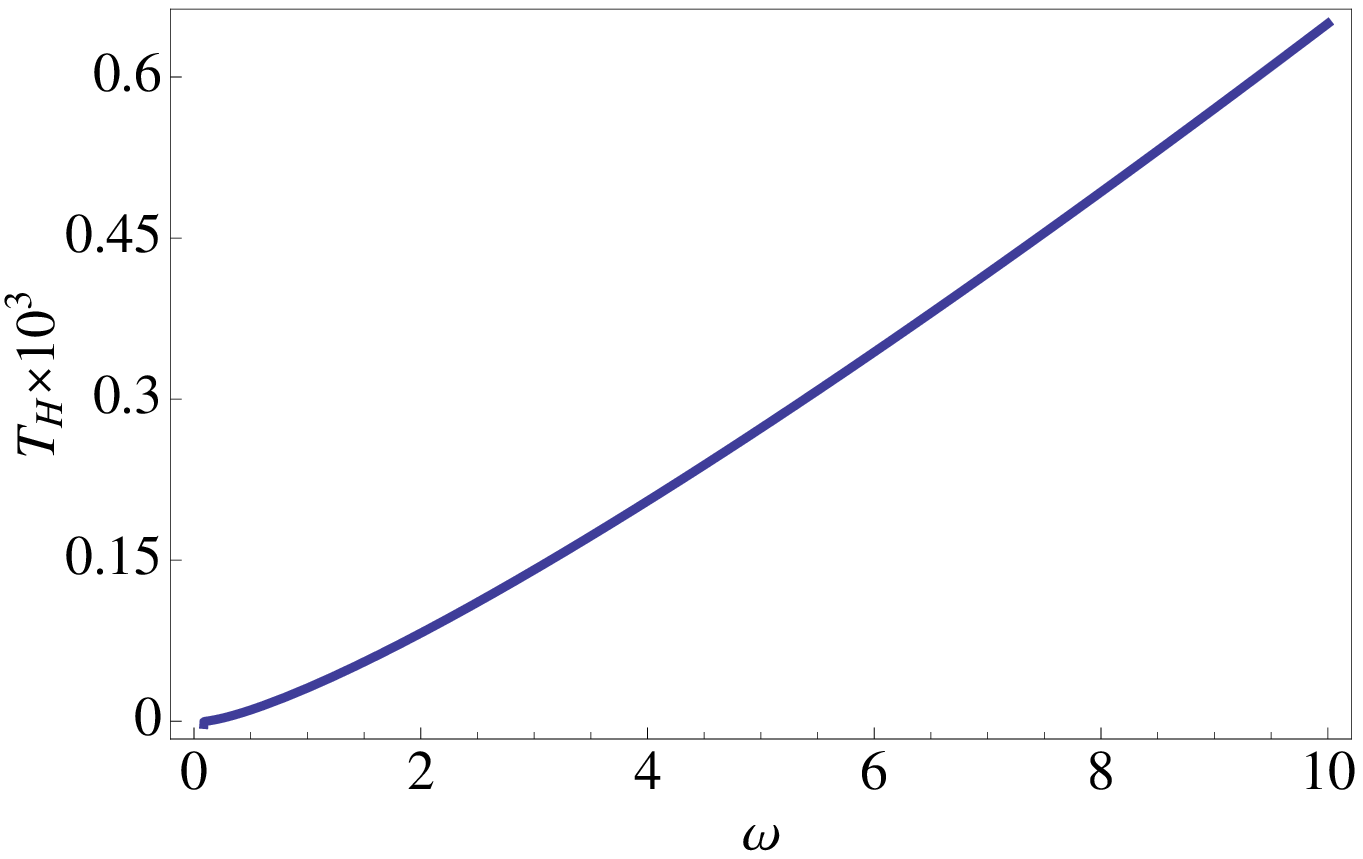,width=.50\textwidth}\\
\epsfig{file=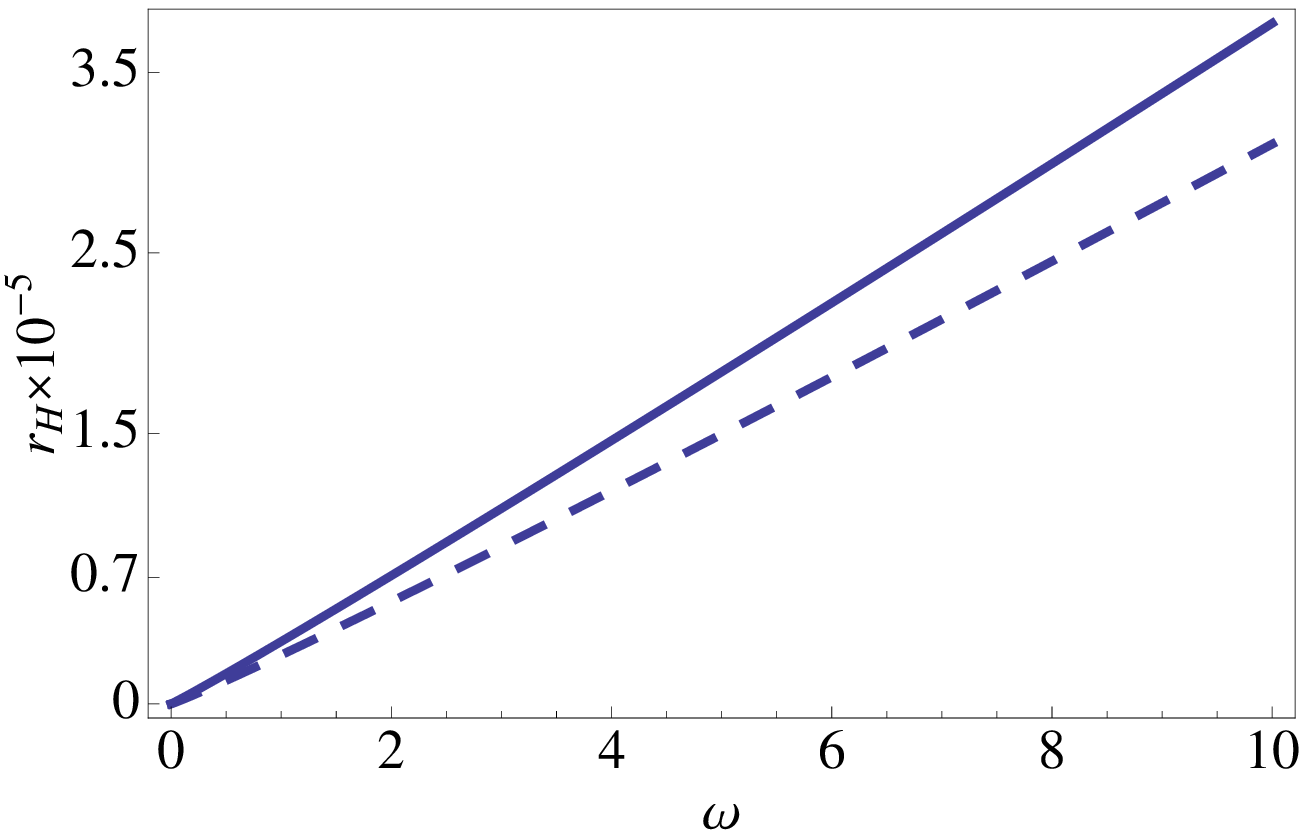,width=.50\textwidth}~~\nobreak
\epsfig{file=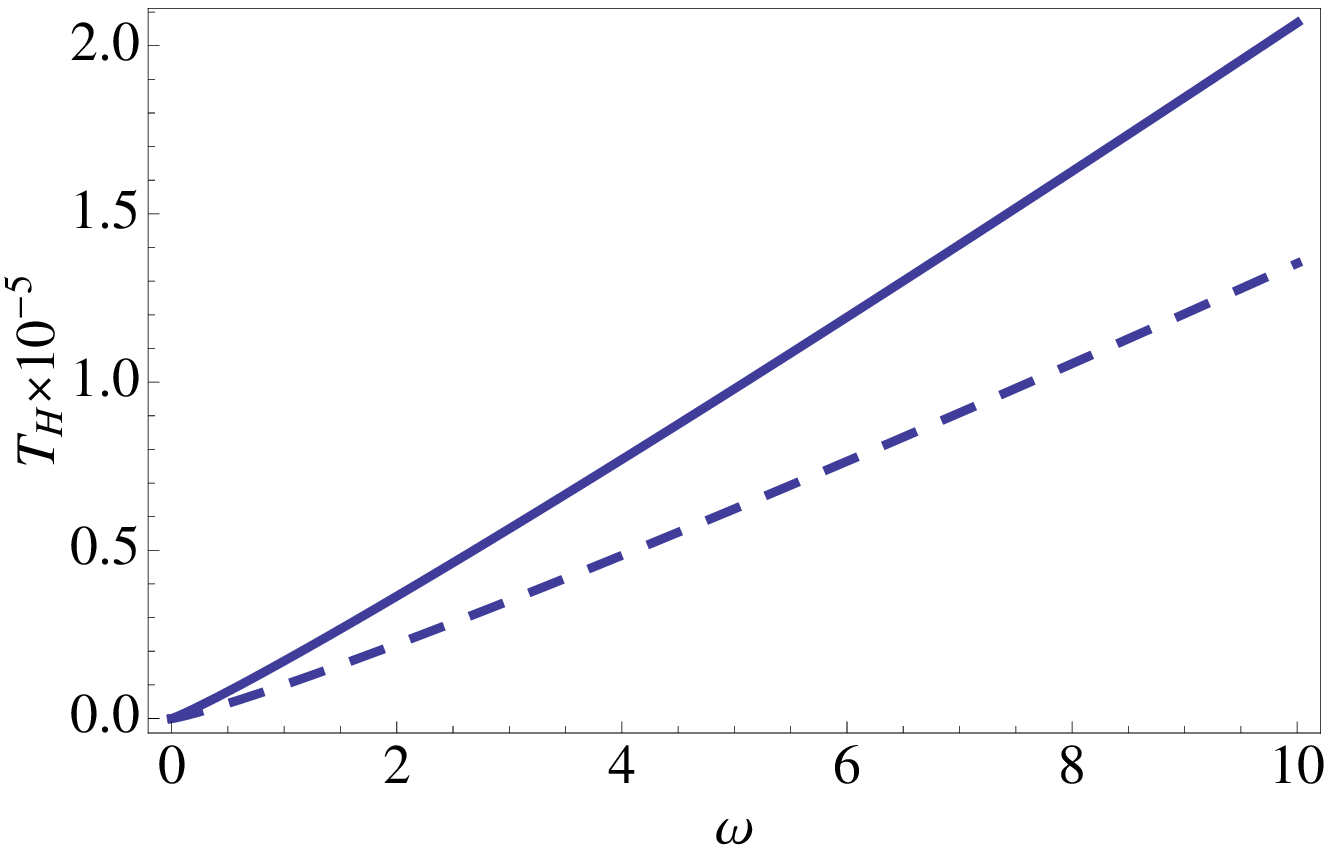,width=.50\textwidth}
\caption{[Upper panel] The behaviour of the
horizon and temperature for $L^2=10^{-2}$, $\epsilon=10^{-5}$
(solid), $\epsilon=10^{-3}$ (dotted), $\epsilon=5\times 10^{-3}$
(dashed). [Lower panel] The behaviour of the 
horizon and temperature for $L^2=10^{2}$, $\epsilon= 10^{-5}$
(solid), $\epsilon=10^{-2}$ (dashed).}
\label{fig:KTps}
\end{center}
\end{figure}

It is also instructive to inspect the parameter dependence of the world volume black hole
solution more closely and quantitatively. Inspection of (\ref{KTtem.}) and (\ref{KThor.})
shows that for the reasonable choice of parameters $L^2=10^{-2}$ and $\epsilon=10^{-5}$
the horizon and temperature of the world volume black hole increase continuously and 
monotonically with increasing the angular velocity (see Fig.\,\ref{fig:KTps} [upper panel]),
as they should. However, inspection of (\ref{KThor.}) shows that by increasing the value of
$\epsilon$, while keeping $L^2$ fixed small, as before, the world volume horizon develops
three branches. It first increases,  then decreases, and finally it increases continuously
(see again Fig.\,\ref{fig:KTps} [upper panel]). Due to the behavior of the world volume
horizon in the middle  branch (where it decreases despite increasing the angular velocity), 
it seems that only for certain values of $\epsilon$ the world volume horizon behaves consistently,
i.e., increases continuously with increasing angular velocities. Nonetheless, inspection of
(\ref{KTtem.}) and (\ref{KThor.}) shows that when the value of $L^2$ is also increased, the
world volume horizon and temperature increase continuously with increasing the angular
velocities despite increasing the value of $\epsilon$ (see Fig.\,\ref{fig:KTps} [lower panel]). 
It is also interesting to see that in this case increasing the value of $\epsilon$ decreases
the world volume horizon and temperature. We therefore conclude that the scale and behavior
of the world volume horizon and temperature in KT depends on the choice of parameters. We
conclude, in particular, that world volume horizons and temperatures of expected features
form in the KT throat subject to certain hierarchies of scales, $\epsilon/g_sM\simeq 10^3-10^4$.

We also note that for $g_sM\ll 1$ our Hawking temperature (\ref{KTtem.}) takes
the general form  $T_{\text{H}}\simeq\hat{r}_{\text{H}}^3/L^2
\sqrt{\ln\left(\hat{r}_{\text{H}}/\epsilon^{2/3}\right)}$. The denominator of this
equals the denominator of the Hawking temperature discussed in \cite{Buchel:2000ch}
(\emph{cf}. Eq.\,(89) of \cite{Herzog:2001xk}), $T_{\text{H}}\simeq\hat{r}_{\text{H}}/L^2
\sqrt{\ln\left(\hat{r}_{\text{H}}/\epsilon^{2/3}\right)}$, but its numerator is very
different. Here it grows with the  cube of ${\hat{r}_{\text{H}}}$ whereas there
(\emph{cf}. Eq.\,(89) of \cite{Herzog:2001xk}) the  numerator of $T_H$ grows linearly
with $\hat{r}_H$. Thus these temperatures, though similar, are not the same. Remember
that the black hole here  lives on a rotating probe brane whereas in \cite{Buchel:2000ch}
there are no probes and the black hole lives in the supergravity background itself. Note
also that here we considered the pure, original KT supergravity solution, neither modifying
it nor considering backreaction effects, as in the probe limit. But, in \cite{Buchel:2000ch}
the KT solution is modified and generalized where it was shown that at sufficiently high
temperature the system develops a horizon with a corresponding Hawking temperature. It is
interesting to note that here we have found that the induced metric on the rotating probe
in KT has thermal horizon and Hawking temperature even when there is no such black hole
in the bulk.

In our analysis above, the backreaction of the D1-brane to the supergravity background solution
has been neglected since we considered the probe limit. It is also instructive to see to what
extend this can be justified. To see this, we note that the total energy of our rotating
D1-brane in the KT supergravity solution is given by the following relation\footnote{Here we
are using the energy-stress tensor defined by $T_{N}^{M}=\frac{2}{\sqrt{-g}}
\frac{\delta S}{\delta g_{M L}}g_{L N}$, where $g_{MN}$ denotes the bulk metric with $M$ and
$N$ running over all ten coordinates of the ten-dimensional spacetime. This satisfies the 
equation of motion $\nabla_M T_N^M=0$. For static spacetime, this reduces to 
$\partial_M(T_t^M\sqrt{-g})=0$, which leads to the energy $E=\int{dr\sqrt{-g}T_{t}^{t}}$.}:

\begin{equation}
\label{EKT}
E=T_{D1}\int_{r_{IR}}^{r_{UV}}{dr\,\sqrt{-g}T_{t}^{t}}=T_{D1}\int_{r_{IR}}^{r_{UV}}
{d\hat{r}\,\left(1+\frac{\overline{\omega}^2}{2\hat{r}^2}\right)}.
\end{equation}
Since in KT $\epsilon$ is very small (such that $\epsilon^2=0$), from Eq.\,(\ref{EKT})
one can see that the energy density becomes very large, when taking the IR limit
$\hat{r}\rightarrow\epsilon^{2/3}$. In addition, the total angular momentum of our 
rotating D1-brane in KT is given by:

\begin{equation}
\label{LKT}
Q=\frac{\delta S}{\delta\dot{\psi}}=\frac{2 L^4 T_{D1} \overline{\omega}}{3}
\int_{r_{IR}}^{r_{UV}}{\frac{d\hat{r}}{\hat{r}^2}\ln\left(\frac{\hat{r}}
{\epsilon^{2/3}}\right)}.
\end{equation}
Performing the integral in (\ref{LKT}) for $r_{IR}\simeq \epsilon^{2/3}$ 
and $r_{UV}\simeq 10^3 \epsilon^{2/3}$ shows that the total angular momentum
is given by $Q\simeq T_{D1} L^4\overline{\omega}/\epsilon^{2/3}$, which is large
like its energy. Thus in the IR limit the backreaction of the probe D1-brane
to the supergravity background is non-negligible even when $\overline{\omega}$
is not large. It is then reasonable to conclude that the large backreaction
will result in the formation of a black hole in the bulk of KT, centered in the IR.

\section{Temperature in the Klebanov-Witten throat}

\subsection{The Klebanov-Witten solution}

In the absence of $M$ wrapped D5-branes the KT throat solution joins
the Klebanov-Witten throat solution \cite{Klebanov:1998hh}. The KW throat is the 
simplest conifold throat background. The ten-dimensional metric and the self-dual
five-form on the KW throat takes the form \cite{Klebanov:1998hh}

\begin{eqnarray}
\label{10DKWmet}
ds_{10}^2&=&h(\hat{r})^{-1/2}dx_{n}dx^{n}+h(\hat{r})^{1/2}\hat{r}^2
\left[\frac{d\hat{r}^2}{\hat{r}^2}+\frac{1}{9}(g^5)^2+\frac{1}{6}
\sum_{i=1}^{4}(g^i)^2\right],\\ \label{F5KW} \tilde{F}_{5}&=&dC_4=
(1+\star_{10})\left[d(h^{-1}(\hat{r}))\wedge dx^0\wedge dx^1
\wedge dx^2 \wedge dx^3\right],
\end{eqnarray}
where the warp factor reads

\begin{equation}
h(\hat{r})=\frac{L^4}{\hat{r}^4},\;\;\;\ \text{and}\;\;\;\ L^4\equiv \frac{27\pi}{4}
g_sN(\alpha^{\prime})^2.
\end{equation}
Note that the asymptotic UV metric, (\ref{10DKWmet}),  has a $U(1)$ symmetry associated
with the rotations of the angular coordinate $\beta=\psi/2$, normalized such that $\beta$
has period $2\pi$. This is the R--symmetry of the dual gauge theory. In the absence of
fractional branes there are no background three-form fluxes, so the $U(1)$ R--symmetry is
a true symmetry of the field theory. Because the R--symmetry is realized geometrically by
invariance under a rigid shift of the angle $\beta$, it becomes a local symmetry in the full
gravity theory, and the associated gauge fields $A=A_{\mu}dx^{\mu}$ appear as fluctuations
of the ten-dimensional metric and RR four-form potential \citep{Ceresole:1999zs,Kim:1985ez,
Gunaydin:1984fk}. The metric and the self-dual five form take modified the form
\cite{Herzog:2001xk}:

\begin{eqnarray}
\label{10DKWmet2}
ds_{10}^2&=&h(\hat{r})^{-1/2}dx_{n}dx^{n}+h(\hat{r})^{1/2}\hat{r}^2
\left[\frac{d\hat{r}^2}{\hat{r}^2}+\frac{1}{9}(g^5-2A)^2+\frac{1}{6}
\sum_{i=1}^{4}(g^i)^2\right]\\ \label{F5KW2} \tilde{F}_5&=& dC_4=
\frac{1}{g_s}d^4x\wedge dh^{-1}+\frac{\pi\alpha^{\prime}N}{4}\Big[(g^5-2A)
\wedge g^1\wedge g^2\wedge g^3\wedge g^4\notag\\ &&-dA\wedge g^5\wedge 
dg^5+\frac{3}{L}\star_{5}dA\wedge dg^5\Big].
\end{eqnarray}
Here note that this metric is of the familiar Kaluza-Klein form, and $\star_{5}$ is the
five-dimensional Hodge star operator defined with respect to the $AdS_5$ metric 
$ds_5^2=h^{-1/2}dx_n\,dx^n+h^{1/2}dr^2$. The supergravity field equation $d\tilde{F}_5=0$
implies \cite{Herzog:2001xk}:

\begin{equation}
\label{Aeq}
d\,\star_5dA=0.
\end{equation}
Hence the $A$ field satisfies the equation of motion for a massless vector in $AdS_5$ space.

\subsection{Induced metric and Hawking temperature in the KW throat}

In this section we derive the world volume horizon and temperature
of the rotating probe in the background (\ref{10DKWmet})--(\ref{F5KW}),
where the background gauge field $A$ is not activated.
Considering the same $S^3$ cycle as in previous sections, we obtain
this full background metric on the KT throat in the form

\begin{equation}
\label{10KWUV}
ds_{10}^2=\frac{\hat{r}^2}{L^2}\left(dx^2-dt^2\right)+\frac{L^2}
{\hat{r}^2}\left(d\hat{r}^2+\frac{\hat{r}^2}{6}d\phi^2+\frac{\hat{r}^2}{9}d\psi^2\right).
\end{equation}

To evaluate the action of the brane, we consider the same embedding ansatz
as in the previous sections with $\phi(\hat{r},t)$ and/or $\psi(\hat{r},t)$ 
specifying the world volume of the brane. To Evaluate (\ref{DPACTION})
in the background (\ref{10KWUV}) (where $M=0$ and $g^{2,4}=0$, giving $C_2=B_2=0$)
for the embedding ansatz, it suffices to set either $\phi$ or $\psi$ constant.
Setting $\phi=const.$, gives the action of the rotating D1-brane in the form:

\begin{eqnarray}
\label{DBIACKW}
S_{\text{D1}}&=&-g_s\,T_{D1}\int{dt\,d\hat{r}\sqrt{1+\frac{\hat{r}^2(\psi^{\prime})^2}{9}
-\frac{L^4\,\dot{\psi}^2}{9\,\hat{r}^2}}}.
\end{eqnarray}
As in the KT and KS throats, we note that in the limit of small velocities
the higher-order non-canonical kinetic terms in (\ref{DBIACKW}) may be dropped.
The Lagrangian and equation of motion for the
slow rotating D1-brane take the simple form:

\begin{equation}
L=1+\frac{\hat{r}^2(\psi^{\prime})^2}{18}
-\frac{L^4\,\dot{\psi}^2}{18\,\hat{r}^2},
\end{equation}

\begin{eqnarray}
\label{BEqnoA1}
\frac{\partial}{\partial\hat{r}}\left[\frac{\hat{r}^2\psi^{\prime}
(\hat{r},t)}{9}\right]&=&\frac{\partial}{\partial t}\left[\frac{L^4\,
\dot{\psi}(\hat{r},t)}{9\hat{r}^2}\right].
\end{eqnarray}

As before, consider solutions of the form

\begin{eqnarray}
\label{RoSolnoA1}
\psi(\hat{r},t)=\omega\,t-\frac{\omega}{\hat{r}}+\psi_0.
\end{eqnarray}
Putting these into the background, gives the induced metric on the brane as

\begin{eqnarray}
\label{indmet1KW}
ds_{ind}^2&=& -\frac{\left[\hat{r}^2-L^4\,\overline{\omega}^2\right]}{L^2}
dt^2\notag+L^2\left(\frac{1}{\hat{r}^2}+\frac{\overline{\omega}^2}{\hat{r}^4}
\right)d\hat{r}^2 + \frac{2\overline{\omega}^2}{\hat{r}^2}L^2dt\,d\hat{r},
\;\;\;\ \overline{\omega}^2=\frac{\omega^2}{9}.\notag\\
\end{eqnarray}
To eliminate the cross term in this metric, we may
consider a coordinate transformation of the form

\begin{equation}
\tau=t-L^4\overline{\omega}^2\int{\frac{d\hat{r}}{\hat{r}^2(\hat{r}^2
-L^4\overline{\omega}^2)}}.
\end{equation}
The induced metric (\ref{indmet1KW}) then takes the form

\begin{eqnarray}
\label{indmet2KW}
ds_{ind}^2&=& -\frac{\left[\hat{r}^2-L^4\,\overline{\omega}^2\right]}{L^2}
d\tau^2+L^2\left[\frac{\overline{\omega}^2+\hat{r}^2-L^4\,
\overline{\omega}^2}{\hat{r}^2(\hat{r}^2-L^4\,\overline{\omega}^2)}\right]d\hat{r}^2.
\end{eqnarray}

It is interesting to note that (for $L=1$) the induced metric, (\ref{indmet2KW}), 
has the form of the BTZ black hole with the angular coordinate suppressed. Thus
it has a world volume horizon and Hawking temperature given by:

\begin{eqnarray}
\label{KWhoreq.}
\hat{r}_H^2-L^8\overline{\omega}^2&=&0,\;\;\;\;\;\;
T_{\text{H}}=\frac{\hat{r}_H}{2\pi}=\frac{L^4\overline{\omega}}{2\pi}.
\end{eqnarray}
Clearly, $\hat{r}_H$ in (\ref{KWhoreq.}) has one (real positive) zero, forming a single
horizon, $\hat{r}_H=L^2\overline{\omega}$. Since the validity range of the solution is
the same as in KT, $\hat{r}_H$ and $T_H$, (\ref{KWhoreq.}), in KW are constrained by
UV/IR scales of the throat. Comparing (\ref{KWhoreq.}) to (\ref{KThor.}) shows that
$\hat{r}_H$ in KW has a form similar to $\hat{r}_H$ in KT for very small rotations,
but shrinking/expanding linearly with $\overline{\omega}$, since there is no logarithmic
warping, and therefore changing more rapidly. This implies that  $T_H$ in KW 
increases/decreases faster than $T_H$ in KT. Note also from the previous section that
in the IR, $\hat{r}_H\rightarrow \epsilon^{2/3}$, and UV, 
$\hat{r}_H\rightarrow 10^2\epsilon^{2/3}$, limits the world volume temperature in KT is
more or less constant and given by $T_H\sim L^2\,\epsilon^{2/3}$. (Also recall from the
previous section that in the IR limit of the KT solution the flux has to be $L^2\gg 1$,
in order to have a valid SUGRA solution, whereas in the UV limit, at sufficiently large
radii, the solution remains valid even if $L^2\ll 1$). On contrary, (\ref{KWhoreq.})
shows $T_H$ in KW linearly increases/decreases with $\hat{r}_H$ in these limits. In
addition to these qualitative differences, there are also quantitative differences in
the two examples in their UV/IR limits.  For $\hat{r}_H\rightarrow \epsilon^{2/3}$, $T_H$
in KW is about $L^2$ times less than $T_H$ in KT, noting that in the IR limit of KT
$L^2\gg 1$. For $\hat{r}_H\rightarrow (10)^2 \epsilon^{2/3}$, $T_H$ in KW is less than
$T_H$ in KT if $L^2\gg 1$ is large enough  whereas if $L^2\ll 1$ $T_H$ in KW is always
greater than $T_H$ in KT, as in this case $T_H$ in KT is vanishingly small. It is 
conclusive then that in the IR limit there is always a large separation between $T_H$
in KW and $T_H$ in KT, by the addition of flux in KT, whereas in the UV limit the separation
may persist, or go away, depending on the choice of flux in KT. It is also clear from 
Eq.\,(\ref{KWhoreq.}) that for rotating probe D1-branes in KW, $\hat{r}_H$ and $T_H$ have
a form very similar to $\hat{r}_H$ and $T_H$ for rotating D1 probes in  $AdS_5\otimes S^5$,
\cite{Das:2010yw}, in particular, increasing/decreasing linearly with $\overline{\omega}$. 

If one considers the backreaction of the above solution to the KW supergravity background, it
is natural to expect such D1-branes to form  mini black holes in the bulk KW. This indicates
that the rotating D1-brane describes  thermal object with temperature $T_H$ in the dual field
theory. The configuration is dual to $\mathcal{N}=1$ conformal gauge theory coupled to a 
monopole at finite temperature. The $\mathcal{N}=1$ gauge theory is itself at zero temperature
while the monopole is at finite temperature $T_H$. Thus such configurations are in non-equilibrium
steady states.

In our analysis above, the backreaction of the D1-brane to the supergravity background
has been neglected since we considered the probe limit. It is also instructive to see
to what extend this can be justified. We note that the total energy of our rotating 
D1-brane in KW is given by:
\begin{equation}
\label{EKW}
E=T_{D1}\int{dr\,\sqrt{-g}T_{t}^{t}}=T_{D1}\int{d\hat{r}\,\left(1+\frac{\overline
{\omega}^2}{2\hat{r}^2}\right)}.
\end{equation}
From Eq.\,(\ref{EKW}) one can see that in the IR the energy density becomes very large,
when taking the IR limit. Furthermore, the total angular momentum of our rotating D1-brane
in KW reads:

\begin{equation}
\label{LKW}
Q=\frac{\delta S}{\delta\dot{\psi}}=\frac{2 L^4 T_{D1} \overline{\omega}}{3}
\int{\frac{d\hat{r}}{\hat{r}^2}}.
\end{equation}
From Eq.\,(\ref{LKW}) one can show that the total angular momentum is large like its energy.
Thus in the IR limit the backreaction of the probe D1-brane to the supergravity background
is non-negligible even when $\overline{\omega}$ is not large. It is then reasonable to
conclude that the large backreaction will result in the formation of a black hole in the bulk
of KW, centered at the IR location.

Now we would like to derive the world volume horizon and temperature of the rotating probe in the background
(\ref{10DKWmet2})--(\ref{F5KW2}), where the background gauge field $A$ is activated and described by (\ref{Aeq}). 
To derive $A$ from Eq.\,(\ref{Aeq}), we note that the four noncompact spatial coordinates combine with the radial
coordinate of the conifold to span $AdS_5$ with its metric given by $ds_5^2=h^{-1/2}dx_n\,dx^n+h^{1/2}dr^2$.
Using this metric, and choosing the gauge $\mathcal{A}^{\mathbf{x}}=\mathcal{A}^r=0$ and $\mathcal{A}^t=\Phi(r)$
one has componentwise as $A_i=g_{ij}\mathcal{A}^j=(h^{-1/2}\mathcal{A}^t, h^{-1/2}\mathcal{A}^{\mathbf{x}},h^{1/2}
\mathcal{A}^{r})=(\widetilde{\Phi}(r),0,0,0)$, where $h=L^4/r^4$. A straightforward but tedious computation shows
that for the above gauge configuration the supergravity equation of motion, (\ref{Aeq}), takes the form:

\begin{eqnarray}
\label{Aeq}
d\star_5\,dA &=& \frac{3!}{L^3}\partial_{r}\big[r^3\,\partial_r
\widetilde{\Phi}(r)\big]dx\wedge dy\wedge dz\wedge dr=0.
\end{eqnarray}
It is easy to see that the solution of (\ref{Aeq}) takes the simple form:

\begin{equation}
\label{tiPhi}
\widetilde{\Phi}(r)\simeq \frac{1}{r^2}.
\end{equation}
We would like to remark that our solution (\ref{tiPhi}) is of expected form.
We note that an electric field will be supported by a potential of the from
$\widetilde{\Phi}(r)=A_t\simeq r^{-2}$. As this is a rank one massless field
in $AdS$, it must correspond to a dimension four operator or current in the
gauge theory. This is just what one would expect from an R--current, to which
gauge fields correspond.

In the presence of the above gauge field the background metric (\ref{10DKWmet2})
for our usual $S^3$ round then takes the form:

\begin{eqnarray}
\label{BM2}
ds_{10}^2&=&-\frac{\hat{r}^2}{L^2}\bigg(1-\frac{4L^4\widetilde{\Phi}^2
(\hat{r})}{9\hat{r}^2}\bigg)dt^2
+\frac{L^2}{\hat{r}^2}\bigg(d\hat{r}^2+\frac{\hat{r}^2}{6}d\phi^2+
\frac{\hat{r}^2}{9}d\psi^2
+\frac{4\hat{r}^2}{9}\widetilde{\Phi}(\hat{r})dt\, d\psi \bigg).\notag\\
\end{eqnarray}

As in the pure KW, we may set $\phi=const.$ and obtain the action of the
rotating D1-brane in the background (\ref{BM2}) as:

\begin{eqnarray}
\label{DBIACKW2}
S_{\text{D1}}&=&-g_s\,T_{D1}\int{dt\,d\hat{r}\sqrt{1-\frac{4 L^4\widetilde{\Phi}^2
(\hat{r})}{9\hat{r}^2}+\frac{\hat{r}^2}{9}(\psi^{\prime}(\hat{r},t))^2\bigg(1-
\frac{4L^4\widetilde{\Phi}^2(\hat{r})}{9\hat{r}^2}\bigg)-\frac{L^4\dot{\psi}^2}
{9\hat{r}^2}}}.\notag\\
\end{eqnarray}
As before, we note that in the limit of small velocities the higher-order
non-canonical kinetic terms in (\ref{DBIACKW2}) can be dropped (after Taylor
expansion). The Lagrangian and equation of motion for the slow rotating D1-brane
then take the form:

\begin{equation}
L =\bigg(1-\frac{4L^4\widetilde{\Phi}^2(\hat{r})}{9\hat{r}^2}\bigg)^{1/2}
\left[1+\frac{\hat{r}^2\,(\psi^{\prime}(\hat{r},t))^2}{18}
-\frac{L^4\dot{\psi}^2}{18\hat{r}^2}\bigg(1-\frac{4L^4\widetilde{\Phi}^2(\hat{r})}
{9\hat{r}^2}\bigg)^{-1}\right],
\end{equation}

\begin{eqnarray}
\label{BEqA2}
\frac{\partial}{\partial\hat{r}}
\left[\bigg(1-\frac{4L^4\widetilde{\Phi}^2(\hat{r})}{9\hat{r}^2}\bigg)^{\frac{1}{2}}
\frac{\hat{r}^2\psi^{\prime}(\hat{r},t)}{9}\right]&=&\frac{\partial}
{\partial t}\left[\bigg(1-\frac{4L^4\widetilde{\Phi}^2(\hat{r})}{9\hat{r}^2}
\bigg)^{-\frac{1}{2}}\frac{L^4\dot{\psi}(\hat{r},t)}{9\hat{r}^2}\right].\;\;\;
\end{eqnarray}
Consider the rotating solution of the form:

\begin{eqnarray}
\label{RoSolA2}
\psi(\hat{r},t)&=&\omega t+u(\hat{r}),\;\;\ u^{\prime}(\hat{r})=\frac{\omega}
{\hat{r}^2(1-4L^4\widetilde{\Phi}^2(\hat{r})/9\hat{r}^2)^{1/2}}.
\end{eqnarray}
Inspection of Eqs.\,(\ref{RoSolA2}) and (\ref{BEqA2}) shows
that in the absence of background gauge field the brane equation of motion
and rotating brane solution reduce to Eqs.\,(\ref{RoSolnoA1})
and (\ref{BEqnoA1}), respectively, as they should.

Putting the solution (\ref{RoSolA2}) into the background metric
(\ref{BM2}), gives the induced metric in the form:

\begin{eqnarray}
\label{indmetA1KW}
ds_{ind.}^2&=&-\frac{(\hat{r}^2-L^4
\overline{\omega}^2-4L^4 \widetilde{\Phi}(\hat{r})(\widetilde{\Phi}(\hat{r})
+\omega)/9)}{L^2}dt^2 \notag\\ &&+L^2 \left[\frac{\overline{\omega}^2}
{\hat{r}^4(1-4L^4\widetilde{\Phi}(\hat{r})^2/9\hat{r}^2)}+\frac{1}{\hat{r}^2}
\right]d\hat{r}^2+\frac{2L^2\omega(\omega+2\widetilde{\Phi}(\hat{r}))}{9\hat{r}^2
(1-4L^4\widetilde{\Phi}(\hat{r})^2/9\hat{r}^2)}d\hat{r} dt.\notag\\ 
\end{eqnarray}
To eliminate the cross-term, consider a coordinate transformation of the form:

\begin{equation}
\tau=t-\frac{L^4 \omega}{9}\int{\frac{d\hat{r}\,(\omega+2\tilde{\Phi}(\hat{r}))}
{\hat{r}^2(1-4L^4 \widetilde{\Phi}^2(\hat{r})/9\hat{r}^2)^{1/2}(\hat{r}^2-L^4
\overline{\omega}^2-4L^4 \widetilde{\Phi}(\hat{r})(\widetilde{\Phi}(\hat{r})+\omega)/9)}}.
\end{equation}
The induced metric then takes the form:

\begin{eqnarray}
\label{indmetA2KW}
ds_{ind.}^2&=&-\frac{\mathcal{G}(\hat{r})}{L^2}d\tau^2\notag\\ &&
+L^2\left[\frac{[9\hat{r}^2-4L^4\widetilde{\Phi}^2(\hat{r})+\omega^2]
\mathcal{G}(\hat{r})+L^4\overline{\omega}^2[\omega
+2\widetilde{\Phi}(\hat{r})]^2}{\hat{r}^2[9\hat{r}^2-4L^4 \widetilde{\Phi}^2
(\hat{r})]\mathcal{G}(\hat{r})}\right]d\hat{r}^2.\notag\\
\end{eqnarray}
Here $\mathcal{G}(\hat{r})=\hat{r}^2-L^4
\overline{\omega}^2-4L^4 \widetilde{\Phi}(\hat{r})(\widetilde{\Phi}(\hat{r})+\omega)/9$. It
is clear that the induced metric depends on the gauge field component $\tilde{\Phi}(\hat{r})$,
given by (\ref{tiPhi}). Inspection of Eqs.\,(\ref{indmetA1KW})--(\ref{indmetA2KW}) shows that
in the absence of the background gauge field the induced world volume metrics reduce to 
(\ref{indmet1KW})--(\ref{indmet2KW}), with the induced world volume horizons and temperatures
given by (\ref{KWhoreq.}). According to our solution (\ref{tiPhi}), this case corresponds to
the very large radii limit, $\hat{r}\rightarrow \infty$.

\begin{figure}[!ht]
\begin{center}
\epsfig{file=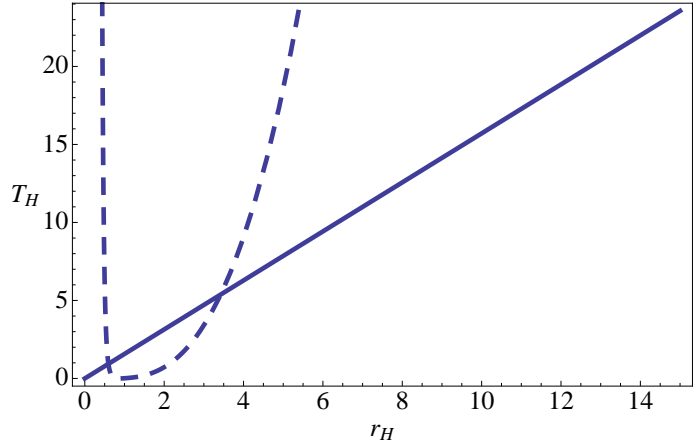,width=.6\textwidth}
\caption{The behaviour of the world volume Hawking temperature for $L=1$ with (dashed)
and without (solid) background gauge field turned on.}
\label{fig:KWp1} 
\end{center}
\end{figure}

Now we would like to take the small radii limit, $\hat{r}\rightarrow 0$, where the
conifold singularity is approached, where massless degrees of freedom become important
(\emph{cf}. \cite{Polchinski:1998rr}). In this limit, our solution (\ref{tiPhi}) shows
that the additional massless gauge field, $\tilde{\Phi}(\hat{r})$, contributes nontrivially
to induced world volume metric, given by the black hole geometry, (\ref{indmetA2KW}). 
The world volume horizon from (\ref{indmetA2KW}),
$\mathcal{G}(\hat{r}_H)=-g_{\tau\tau}=g^{\hat{r}\hat{r}}=0$, is described by: 
\begin{eqnarray}
\hat{r}_{H}^2-L^4\overline{\omega}^2-4L^4 \widetilde{\Phi}(\hat{r}_H)(\widetilde{\Phi}
(\hat{r}_H)+\omega)/9&=& 0,\notag\\ \label{KWHorAeq}
\hat{r}_H^6-L^4\overline{\omega}^2\hat{r}_H^4-(4L^4\omega/9)\hat{r}_H^2-4L^4/9&=&0.
\end{eqnarray}
Here in the second line we inserted the gauge field component given by our solution (\ref{tiPhi}).
It is clear from (\ref{KWHorAeq}) that near the conifold point the massless gauge field contributes
nontrivially to the world volume horizon. It is interesting to note that (\ref{KWHorAeq}) has the
form of the horizon equation of the $AdS$--Reissner-Nordstr\"om black hole 
(\emph{cf}. \cite{Johnson:2003gi}). Note though, as before, that here the horizon (\ref{KWHorAeq})
is on the world volume obtained from the induced metric, (\ref{indmetA2KW}), on the rotating a probe
in the KW, without having a black hole in the KW background. Equation (\ref{KWHorAeq}) is a `sextic
equation' in $\hat{r}_H$. To solve this equation, we reduced it to a `depressed cubic' and use
trigonometric method to find the three distinct roots. However, we note that the horizon of the
solution is located at the largest root of $g^{\hat{r}\hat{r}}=0$, which is:

\begin{eqnarray}
\label{HorKTA}
\hat{r}_H^{+} &=&\frac{L^4\overline{\omega}^2}{3}+ \frac{2L^2}{3}
\sqrt{4\omega/3+L^4\overline{\omega}^4}\notag\\ && \times\cos\left\{(1/3)\arccos 
\bigg[3\left(\frac{6+2 L^4\overline{\omega}^2\omega+L^8
\overline{\omega}^6}{4\omega+3L^4\overline{\omega}^4}\right)
\sqrt{\frac{3}{4\omega+3L^4\overline{\omega}^4}}\bigg]\right\}.
\end{eqnarray}

The Hawking temperature at this horizon takes the form:

\begin{eqnarray}
T_H&=&\frac{r^2[9\hat{r}^2-4L^4\widetilde{\Phi}^2(r)][2\hat{r}+4L^4
(2\widetilde{\Phi}(\hat{r})+\omega)\tilde{\Phi}^{\prime}(\hat{r})/9]}
{2\pi L^6 \overline{\omega}^2[\omega+2\widetilde{\Phi}(r)]^2}
\bigg|_{\hat{r}=\hat{r}_H^+}
\notag\\  \label{THKTA} &=&
\frac{r^2[9(\hat{r}_H^+)^2-4L^4/(\hat{r}_H^+)^4][2\hat{r}_H^{+}
-4L^4(2/(\hat{r}_H^+)^2+\omega)/9(\hat{r}_H^+)^3]}
{2\pi L^6 \overline{\omega}^2[\omega+2/(\hat{r}_H^+)^2]^2}.
\end{eqnarray}
Here in the second line we inserted the gauge field component given by our solution (\ref{tiPhi}),
and $\hat{r}_H^+$ is radius of the horizon given by (\ref{HorKTA}). By comparing (\ref{THKTA}) and
(\ref{HorKTA}) with (\ref{KWhoreq.}), it is clear that the presence of the background gauge field
has modified the world volume horizon and temperature significantly. It is also straightforward to
see that by setting the background gauge field zero, $\widetilde{\Phi}(r)=0$, the above temperature
(\ref{THKTA}) reduces to the temperature (\ref{KWhoreq.}), as it should.  Our expression (\ref{THKTA})
is quite interesting and very distinct from our previous world volume temperatures 
(see Fig.\,\ref{fig:KWp1}). This is because for a given temperature $T_H$, there are in fact \emph{two}
values of $\hat{r}_H^+$, which solve the relation (\ref{THKTA}). We also see that there are two classes
of world volume black hole solutions. There is one branch which, for large $\hat{r}_H^+$, the temperature
$T_H$ goes with $(\hat{r}_H^+)^3$. The other branch goes at small $\hat{r}_H^+$ as the inverse of
$(\hat{r}_H^+)^4$. These `small' black holes have the familiar behavior of five-dimensional black holes
in asymptotically flat spacetime, as their temperature decreases with increasing horizon size. (The term
`small' is appropriate, as they are smaller than the characteristic size set by the AdS scale $L$ , and
so they have the characteristics of the asymptotically Minkowskian black holes). In a similar way, the
`large' black holes may be obtained when $L$ is small compared to the horizon size. We also note that
increasing the flux further merely changes the scale of the temperature but leaves its overall behavior
unchanged (as in Fig.\,\ref{fig:KWp1}). We therefore conclude that when in the KW solution the background
gauge field from the $U(1)$ R--symmetry is turned on, the induced metric on the rotating probe in the KW
background admits two classes of world volume black hole solutions. These are characterized by two
qualitatively distinct world volume temperatures in the large and small world volume horizon radii limits,
respectively. This is similar to the behavior of the temperature in $AdS$-Schwarzschild black hole (\emph{cf}.
\cite{Johnson:2003gi}), though note  that the black hole solution we obtained here, as in previous examples,
is on the world volume of the rotating probe in spite of the absence of real black holes in the bulk.

In our analysis above, the backreaction of the D1-brane to the supergravity background has been neglected
since we considered the probe limit. It is also instructive to see to what extend this can be justified.
We note that the total energy of our rotating D1-brane in KW with background gauge field turned on is given by:
\begin{eqnarray}
\label{EKW2}
E&=&T_{D1}\int{dr\,\sqrt{-g}T_{t}^{t}}\notag\\ &=&2T_{D1}\int{d\hat{r}
\bigg(1-\frac{4L^4 \widetilde{\Phi}^2(r)}{9\hat{r}^2}\bigg)^{1/2}}
\bigg[1+\frac{\overline{\omega}^2}{2 \hat{r}^2(1-4L^4 \widetilde{\Phi}^2(r)
/9\hat{r}^2)}\bigg]\notag\\ &=&2T_{D1}\int{d\hat{r}(9\hat{r}^6-4L^4)^{1/2}}
\bigg[1+\frac{\hat{r}^4\omega^2}{2 (9\hat{r}^6-4L^4)}\bigg].
\end{eqnarray}
It is clear that when the background gauge field is turned off, $\widetilde{\Phi}(r)=0$,
Eq.\,(\ref{EKW2}) reduces to Eq.\,(\ref{EKW}), as it should.
From Eq.\,(\ref{EKW2}) one can see that the energy density becomes very large,
when taking the IR limit. Furthermore, the total angular momentum of our rotating D1-brane
in KW reads:

\begin{equation}
\label{LKW2}
Q=\frac{\delta S}{\delta\dot{\psi}}=T_{D1} \frac{L^4\omega}{3}
\int{\frac{ d\hat{r}}{\hat{r}^2[1-4L^4 \widetilde{\Phi}^2(r)/9\hat{r}^2]^{1/2}}}=T_{D1}
\frac{L^4\omega}{3}\int{\frac{\hat{r} d\hat{r}}{(9\hat{r}^6-4L^4)^{1/2}}}.
\end{equation}
It is clear that when the background gauge field is turned off, $\widetilde{\Phi}(r)=0$,
Eq.\,(\ref{LKW2}) reduces to Eq.\,(\ref{LKW}), as it should. From Eq.\,(\ref{LKW2}) one can
see that in the IR the angular momentum is large like its energy. Thus in the IR limit the
backreaction of the probe D1-brane to the supergravity background is non-negligible even when
$\omega$ is not large. It is then reasonable to conclude that the large backreaction will result
in the formation of a black hole in the bulk of KW, centered in the IR region.

\section{Discussion}

In this paper, we studied the induced world volume metrics on rotating probes in warped
Calabi-Yau throats and provided the first examples of world volume black hole solutions
in such throat backgrounds. Such examples have been found in the literature in 
$AdS_5\otimes S^5$ background dual to $\mathcal{N}=4$ SYM theory. The aim of our work
was to extend such examples to more general supergravity solutions including Calabi-Yau
throat backgrounds dual to $\mathcal{N}=1$ gauge theories. The motivation of our study
was to find examples of world volume black hole solutions in backgrounds where some
supersymmetry and/or conformal invariance are broken whereby the IR behavior of $AdS$
spacetime is modified. By gauge/gravity duality, the temperature of such black hole 
solutions correspond to the temperatures of flavors in the dual gauge field theory. We
considered different examples of conifold throat backgrounds and computed the induced
metrics on the world volume of rotating probe branes in these backgrounds. We performed
a UV/IR consistent analysis and found interesting novel black hole solutions that can
form on the world volume of the rotating probe brane in these backgrounds in spite of
the absence of black holes in the bulk Calabi-Yau.

We began by considering the KS solution as the first example, taking its very small
radii limit corresponding to the very deep IR limit of the solution. We then modified
the background by taking the very large radii limit corresponding to UV solutions,
considering respectively the KT and KW solutions as next examples. Expectedly, the
world volume horizons and temperatures in these two UV solutions present some
similarities, due to the fact that warping has similar features in these solutions.
Nonetheless, comparing these, in particular, with the analysis in the deep IR limit
of the KS,  our study revealed interesting differences. These differences are related
both to the varying forms of the warp factors in the three cases, and to the different
manifolds in which the angular coordinates are compactified. We also found interesting
differences between world volume horizons and temperatures in KW, when the background
gauge field due to $U(1)$ R--symmetry is activated.

In the very deep IR limit of the KS solution, where the warp factor is constant
and the solution is regular, whereby the dual field theory is confining and breaking
chiral symmetry, we derived the induced world volume metric on the rotating probe and
found no world volume black hole nucleation with horizons and temperatures of expected
features. On contrary, in the UV solution, including the KT throat, where the warp
factor varies logarithmically and the solution is singular, whereby the dual field theory
is non-confining and chiral, we found from induced world volume metric on the rotating
probe world volume black hole solutions with horizons and temperatures of expected
features. In the UV solution, including the KW throat, where the warp factor is that of
$AdS$ and the dual field theory is conformal, we also found that the induced world volume
metric on the probe is given by the black hole geometry with horizons and temperatures of
expected features.

In both KT and KW examples, we found that the induced world volume metrics on the rotating
probes have single world volume horizons and finite world volume temperatures. In the KT
solution, we found that world volume horizon on the rotating probe is given by the Lambert's
transcendental equation solving to the Lambert function. We found that the world volume horizon
forms about the singularity with the horizon size growing with increasing the angular momenta.
We found that certain angular momenta guarantee the world volume horizons and temperatures to
form in the UV away from the KT singularity in the IR. Taking the limits of the world volume
temperature, as the world volume horizon approaches the UV/IR regions of the throat, we found
that due to logarithmic warping the world volume temperature of the rotating probe in KT is
more or less constant and determined by the flux: For large enough flux and world volume 
horizons approaching the UV/IR, we found that the world volume temperatures of the slow rotating
probe are finitely small. On contrary, for small flux and world volume horizons approaching the
UV we found that the world volume temperature of the slow rotating probe is vanishingly small.
Hence we found that the size of the world volume temperature in KT in the far UV limit depends
on the choice of flux and accordingly compares to the size of the world volume temperature in
the IR limit. In KT we also examined the parameter dependence of the solution and found that
the scale and behavior of the world volume horizons and temperatures are subject to certain
hierarchies between flux and the deformation parameter. In KW, where logarithmic warping is
removed, we found that the induced world volume metric on the rotating probe has the form of
the BTZ black hole. We found the world volume horizons and temperatures varying linearly with
angular velocities. These were  different from world volume horizons and temperatures of rotating
probes in KT, but similar to those of rotating probes in $AdS_5\otimes S^5$ found in the 
literature. Taking the limits of world volume temperatures, as the world volume horizon approaches
the UV/IR, we found that in the IR limit the world volume temperatures in KW and in KT are largely
separated by the additional flux in KT whereas in the UV limit we found this difference depending
on the choice of flux in KT. Furthermore, in KW we saw that turning on a background gauge field
due to $U(1)$ R--symmetry modifies the induced world volume metric on the rotating probe. We 
found that the world volume horizons and temperature have features and behaviors similar to those
of $AdS$--Reissner--Nordstr\"om and $AdS$-Schwarzschild black holes. We showed that in this particular
case the related Hawking temperature on the probe has two distinct branches which describe two
different classes of black hole solutions. These included `small' black holes characterized by
temperatures which decrease with increasing horizon size. In both KT and KW, we also discussed
the backreaction of our rotating D1-brane to the SUGRA background. We noted that by taking into
account the backreaction of the rotating brane solution to the SUGRA background the rotating solution
is expected to produce mini black holes in the bulk. We noted that these describe thermal objects in
the dual field theory, including a monopole at finite temperature coupled to the corresponding
$\mathcal{N}=1$ gauge theory which itself is at zero temperature.

We conclude that the induced world volume metrics on slow rotating probe D1-branes in warped
Calabi-Yau throats have thermal horizons with characteristic Hawking temperatures despite the
absence of black holes in the bulk Calabi-Yau. We conclude that this world volume black hole
nucleation depends on the warping and the deformation of the throat with world volume horizons
and temperatures of expected features forming not in the regular confining IR region where
warping is constant, but in the singular UV solutions where warping varies. Furthermore, we
conclude that in the singular UV solutions world volume black hole solutions with distinct
horizons and temperatures form due to distinct choices of warping and background gauge fields. 
By gauge/gravity duality, we may also conclude that the $\mathcal{N}=1$ gauge field theories
at zero temperature couple to thermal objects at finite temperature, hence producing systems
in non-equilibrium steady states, if the supergravity dual containing rotating probes is away
from the confining IR limit.

There are limitations and approximations considered in our work. First, in order to obtain simple
analytic rotating brane solutions, we considered the simplest probe brane, a probe D1-brane, and
took the very small and the very large radii limits of the geometry corresponding to the IR and UV
solutions, respectively, where the explicit analytic form of the warp factors is known. Though
these are the two important limits of the full SUGRA solution, they rather solve the world volume
dynamics in specific conical regions. A more generic analysis for the world volume dynamics would
require the consideration of the full KS warp factor, with the brane equations of motion solved
numerically. Second, in order to keep our analysis of world volume horizon and temperature as simple
as possible, we deliberately turned off gauge fields on the probe D1-brane and considered slow
rotations about cycles inside the throat for which the two-form background fields have a vanishing
contribution to the D1-brane action. This rendered the world volume dynamics of the probe D1-brane
easy to solve. We expect, in particular, that taking other cycles parametrized differently, such
that the contribution of the two-forms to the world volume D1-brane action is locally nontrivial,
to significantly complicate our analysis, and possibly lead to some alterations of our results. 
However, we expect our main conclusions remain unchanged. Third, we did not discuss in this work
the details of thermalization of dual gauge field theory of our results. For instance, we could
consider gauge theory on the probe in the KW theory, compute the related temperature and find 
agreement with the supergravity result for the temperature on the probe discussed here.  The gauge
theory approach can shed light on the temperature for extended objects which we may consider in a
separate work.  Fourth,  the other aspect that we did not study is the connection between the 
Hawking and the Unruh temperatures of the probe brane. One could derive the Hawking temperature of
the rotating probe in supergravity and compare the local temperature with Unruh temperature.

Our analysis in this paper can be extended in several ways. One extension would be to study world
volume horizons and temperatures of higher dimensional rotating branes and make comparison with the
results  obtained in this paper. The other possible extension would be to study horizons and
temperatures on rotating probe branes in the throat subject to moduli stabilization, which induce
corrections to the ISD supergravity solution and hence to the action of the probe brane. Finally,
we note that the complete understanding of the temperature of probes would require the consideration
of the full square root structure of the probe action, including higher order non-canonical kinetic
terms as well as the contribution of world volume gauge fields.  We leave the investigation of
these and the above issues for future study.

\section*{Acknowledgement}

We are grateful to our editor and, especially, to our referee for useful
comments on our paper.


\begin{thebibliography}{99}


\bibitem{Maldacena:1997re} 
  J.~M.~Maldacena,
  ``The Large N limit of superconformal field theories and supergravity,''
  Int.\ J.\ Theor.\ Phys.\  {\bf 38}, 1113 (1999)
  [Adv.\ Theor.\ Math.\ Phys.\  {\bf 2}, 231 (1998)]
  [hep-th/9711200].
  S.~S.~Gubser, I.~R.~Klebanov and A.~M.~Polyakov,
  ``Gauge theory correlators from noncritical string theory,''
  Phys.\ Lett.\ B {\bf 428}, 105 (1998)
  [hep-th/9802109].
  E.~Witten,
  ``Anti-de Sitter space and holography,''
  Adv.\ Theor.\ Math.\ Phys.\  {\bf 2}, 253 (1998)
  [hep-th/9802150].



\bibitem{Gubser:1996de} 
  S.~S.~Gubser, I.~R.~Klebanov and A.~W.~Peet,
  ``Entropy and temperature of black 3-branes,''
  Phys.\ Rev.\ D {\bf 54}, 3915 (1996)
  [hep-th/9602135].
  E.~Witten,
  ``Anti-de Sitter space, thermal phase transition, and confinement in gauge theories,''
  Adv.\ Theor.\ Math.\ Phys.\  {\bf 2}, 505 (1998)
  [hep-th/9803131].







\bibitem{Klebanov:2000hb} 
  I.~R.~Klebanov and M.~J.~Strassler,
  ``Supergravity and a confining gauge theory: Duality cascades and
  chiSB-resolution of naked singularities,''
  JHEP {\bf 0008}, 052 (2000)
  [arXiv:hep-th/0007191].
  


\bibitem{Klebanov:2000nc} 
  I.~R.~Klebanov and A.~A.~Tseytlin,
  ``Gravity duals of supersymmetric $SU(N)\times SU(N+M)$ gauge theories,''
  Nucl.\ Phys.\ B {\bf 578}, 123 (2000)
  [hep-th/0002159].


\bibitem{Klebanov:1998hh} 
  I.~R.~Klebanov and E.~Witten,
  ``Superconformal field theory on three-branes at a Calabi-Yau singularity,''
  Nucl.\ Phys.\ B {\bf 536}, 199 (1998)
  [hep-th/9807080].
  


  
  
\bibitem{Herzog:2001xk} 
  C.~P.~Herzog, I.~R.~Klebanov and P.~Ouyang,
  ``Remarks on the warped deformed conifold,''
  arXiv:hep-th/0108101.
  C.~P.~Herzog, I.~R.~Klebanov and P.~Ouyang,
  ``D-branes on the conifold and N=1 gauge / gravity dualities,''
  hep-th/0205100.
  
\bibitem{Giddings:2001yu} 
  S.~B.~Giddings, S.~Kachru and J.~Polchinski,
  ``Hierarchies from fluxes in string compactifications,''
  Phys.\ Rev.\  D {\bf 66}, 106006 (2002)
  [arXiv:hep-th/0105097].





\bibitem{Buchel:2000ch} 
  A.~Buchel,
  ``Finite temperature resolution of the Klebanov-Tseytlin singularity,''
  Nucl.\ Phys.\ B {\bf 600}, 219 (2001)
  [hep-th/0011146]. 
  A.~Buchel, C.~P.~Herzog, I.~R.~Klebanov, L.~A.~Pando Zayas and A.~A.~Tseytlin,
  ``Nonextremal gravity duals for fractional D-3 branes on the conifold,''
  JHEP {\bf 0104}, 033 (2001)
  [hep-th/0102105].
  S.~S.~Gubser, C.~P.~Herzog, I.~R.~Klebanov and A.~A.~Tseytlin,
  ``Restoration of chiral symmetry: A Supergravity perspective,''
  JHEP {\bf 0105}, 028 (2001)
  [hep-th/0102172].
  
  




  
\bibitem{Karch:2001cw} 
  A.~Karch and L.~Randall,
  ``Localized gravity in string theory,''
  Phys.\ Rev.\ Lett.\  {\bf 87}, 061601 (2001)
  [hep-th/0105108].
  A.~Karch and L.~Randall,
  ``Open and closed string interpretation of SUSY CFT's on branes with boundaries,''
  JHEP {\bf 0106}, 063 (2001)
  [hep-th/0105132].

  
\bibitem{Karch:2002sh} 
  A.~Karch and E.~Katz,
  ``Adding flavor to AdS/CFT,''
  JHEP {\bf 0206}, 043 (2002)
  [hep-th/0205236].
  M.~Kruczenski, D.~Mateos, R.~C.~Myers and D.~J.~Winters,
  ``Meson spectroscopy in AdS / CFT with flavor,''
  JHEP {\bf 0307}, 049 (2003)
  [hep-th/0304032].
  D.~Mateos, R.~C.~Myers and R.~M.~Thomson,
  ``Holographic phase transitions with fundamental matter,''
  Phys.\ Rev.\ Lett.\  {\bf 97}, 091601 (2006)
  [hep-th/0605046].
  D.~Mateos, R.~C.~Myers and R.~M.~Thomson,
  ``Holographic viscosity of fundamental matter,''
  Phys.\ Rev.\ Lett.\  {\bf 98}, 101601 (2007)
  [hep-th/0610184].
  S.~Kobayashi, D.~Mateos, S.~Matsuura, R.~C.~Myers and R.~M.~Thomson,
  ``Holographic phase transitions at finite baryon density,''
  JHEP {\bf 0702}, 016 (2007)
  [hep-th/0611099].
  D.~Mateos, S.~Matsuura, R.~C.~Myers and R.~M.~Thomson,
  ``Holographic phase transitions at finite chemical potential,''
  JHEP {\bf 0711}, 085 (2007)
  [arXiv:0709.1225 [hep-th]].
  J.~Babington, J.~Erdmenger, N.~J.~Evans, Z.~Guralnik and I.~Kirsch,
  ``Chiral symmetry breaking and pions in nonsupersymmetric gauge/gravity duals,''
  Phys.\ Rev.\ D {\bf 69}, 066007 (2004)
  [hep-th/0306018].
  R.~Apreda, J.~Erdmenger, N.~Evans and Z.~Guralnik,
  ``Strong coupling effective Higgs potential and a first order thermal phase transition from AdS/CFT duality,''
  Phys.\ Rev.\ D {\bf 71}, 126002 (2005)
  [hep-th/0504151]].
  G.~Itsios, N.~Jokela and A.~V.~Ramallo,
  ``Collective excitations of massive flavor branes,''
  arXiv:1602.06106 [hep-th].
  
  
\bibitem{Sakai:2003wu} 
  T.~Sakai and J.~Sonnenschein,
  ``Probing flavored mesons of confining gauge theories by supergravity,''
  JHEP {\bf 0309}, 047 (2003)
  [hep-th/0305049]. 
  P.~Ouyang,
  ``Holomorphic D7 branes and flavored N=1 gauge theories,''
  Nucl.\ Phys.\ B {\bf 699}, 207 (2004)
  [hep-th/0311084].
  T.~S.~Levi and P.~Ouyang,
  ``Mesons and flavor on the conifold,''
  Phys.\ Rev.\ D {\bf 76}, 105022 (2007)
  [hep-th/0506021].
  S.~Kuperstein,
  ``Meson spectroscopy from holomorphic probes on the warped deformed conifold,''
  JHEP {\bf 0503}, 014 (2005)
  [hep-th/0411097].
  S.~Kuperstein and J.~Sonnenschein,
  ``A New Holographic Model of Chiral Symmetry Breaking,''
  JHEP {\bf 0809}, 012 (2008)
  [arXiv:0807.2897 [hep-th]].
  O.~Ben-Ami, S.~Kuperstein and J.~Sonnenschein,
  ``On spontaneous breaking of conformal symmetry by probe flavour D-branes,''
  JHEP {\bf 1403}, 045 (2014)
  [arXiv:1310.8366 [hep-th]].
  A.~Dymarsky, S.~Kuperstein and J.~Sonnenschein,
  ``Chiral Symmetry Breaking with non-SUSY D7-branes in ISD backgrounds,''
  JHEP {\bf 0908}, 005 (2009)
  [arXiv:0904.0988 [hep-th]].
  F.~Benini, F.~Canoura, S.~Cremonesi, C.~Nunez and A.~V.~Ramallo,
  ``Unquenched flavors in the Klebanov-Witten model,''
  JHEP {\bf 0702}, 090 (2007)
  [hep-th/0612118].
  F.~Benini, F.~Canoura, S.~Cremonesi, C.~Nunez and A.~V.~Ramallo,
  ``Backreacting flavors in the Klebanov-Strassler background,''
  JHEP {\bf 0709}, 109 (2007)
  [arXiv:0706.1238 [hep-th]].
  F.~Benini,
  ``A Chiral cascade via backreacting D7-branes with flux,''
  JHEP {\bf 0810}, 051 (2008)
  [arXiv:0710.0374 [hep-th]].
  F.~Bigazzi, A.~L.~Cotrone and A.~Paredes,
  ``Klebanov-Witten theory with massive dynamical flavors,''
  JHEP {\bf 0809}, 048 (2008)
  [arXiv:0807.0298 [hep-th]].
  F.~Bigazzi, A.~L.~Cotrone, A.~Paredes and A.~V.~Ramallo,
  ``The Klebanov-Strassler model with massive dynamical flavors,''
  JHEP {\bf 0903}, 153 (2009)
  [arXiv:0812.3399 [hep-th]].
  
  

  

\bibitem{Karch:2007pd} 
  A.~Karch and A.~O'Bannon,
  ``Metallic AdS/CFT,''
  JHEP {\bf 0709}, 024 (2007)
  [arXiv:0705.3870 [hep-th]].
  M.~Ammon, J.~Erdmenger, M.~Kaminski and P.~Kerner,
  ``Superconductivity from gauge/gravity duality with flavor,''
  Phys.\ Lett.\ B {\bf 680}, 516 (2009)
  [arXiv:0810.2316 [hep-th]].
  M.~Ammon, J.~Erdmenger, M.~Kaminski and P.~Kerner,
  ``Flavor Superconductivity from Gauge/Gravity Duality,''
  JHEP {\bf 0910}, 067 (2009)
  [arXiv:0903.1864 [hep-th]].
  N.~Evans, A.~Gebauer, K.~Y.~Kim and M.~Magou,
  ``Holographic Description of the Phase Diagram of a Chiral Symmetry Breaking Gauge Theory,''
  JHEP {\bf 1003}, 132 (2010)
  [arXiv:1002.1885 [hep-th]].
  K.~Jensen, A.~Karch and E.~G.~Thompson,
  ``A Holographic Quantum Critical Point at Finite Magnetic Field and Finite Density,''
  JHEP {\bf 1005}, 015 (2010)
  [arXiv:1002.2447 [hep-th]].
  K.~Jensen, A.~Karch, D.~T.~Son and E.~G.~Thompson,
  ``Holographic Berezinskii-Kosterlitz-Thouless Transitions,''
  Phys.\ Rev.\ Lett.\  {\bf 105}, 041601 (2010)
  [arXiv:1002.3159 [hep-th]].
  N.~Evans, A.~Gebauer, K.~Y.~Kim and M.~Magou,
  ``Phase diagram of the D3/D5 system in a magnetic field and a BKT transition,''
  Phys.\ Lett.\ B {\bf 698}, 91 (2011)
  [arXiv:1003.2694 [hep-th]].
  S.~A.~Hartnoll, J.~Polchinski, E.~Silverstein and D.~Tong,
  ``Towards strange metallic holography,''
  JHEP {\bf 1004}, 120 (2010)
  [arXiv:0912.1061 [hep-th]].
  

  
  

\bibitem{Herzog:2009xv} 
  C.~P.~Herzog,
  ``Lectures on Holographic Superfluidity and Superconductivity,''
  J.\ Phys.\ A {\bf 42}, 343001 (2009)
  [arXiv:0904.1975 [hep-th]].
  S.~A.~Hartnoll,
  ``Lectures on holographic methods for condensed matter physics,''
  Class.\ Quant.\ Grav.\  {\bf 26}, 224002 (2009)
  [arXiv:0903.3246 [hep-th]].
  T.~Faulkner, N.~Iqbal, H.~Liu, J.~McGreevy and D.~Vegh,
  ``From Black Holes to Strange Metals,''
  arXiv:1003.1728 [hep-th].
  J.~McGreevy,
  ``Holographic duality with a view toward many-body physics,''
  Adv.\ High Energy Phys.\  {\bf 2010}, 723105 (2010)
  [arXiv:0909.0518 [hep-th]].
  H.~Liu, J.~McGreevy and D.~Vegh,
  ``Non-Fermi liquids from holography,''
  Phys.\ Rev.\ D {\bf 83}, 065029 (2011)
  [arXiv:0903.2477 [hep-th]].
  T.~Faulkner, H.~Liu, J.~McGreevy and D.~Vegh,
  ``Emergent quantum criticality, Fermi surfaces, and AdS(2),''
  Phys.\ Rev.\ D {\bf 83}, 125002 (2011)
  [arXiv:0907.2694 [hep-th]].



\bibitem{Russo:2008gb} 
  J.~G.~Russo and P.~K.~Townsend,
  ``Accelerating Branes and Brane Temperature,''
  Class.\ Quant.\ Grav.\  {\bf 25}, 175017 (2008)
  [arXiv:0805.3488 [hep-th]].

\bibitem{Chernicoff:2008sa} 
  M.~Chernicoff and A.~Guijosa,
  ``Acceleration, Energy Loss and Screening in Strongly-Coupled Gauge Theories,''
  JHEP {\bf 0806}, 005 (2008)
  [arXiv:0803.3070 [hep-th]].


\bibitem{Paredes:2008cr} 
  A.~Paredes, K.~Peeters and M.~Zamaklar,
  ``Temperature versus acceleration: The Unruh effect for holographic models,''
  JHEP {\bf 0904}, 015 (2009)
  [arXiv:0812.0981 [hep-th]].


\bibitem{Athanasiou:2010pv} 
  C.~Athanasiou, P.~M.~Chesler, H.~Liu, D.~Nickel and K.~Rajagopal,
  ``Synchrotron radiation in strongly coupled conformal field theories,''
  Phys.\ Rev.\ D {\bf 81}, 126001 (2010)
  [Erratum-ibid.\ D {\bf 84}, 069901 (2011)]
  [arXiv:1001.3880 [hep-th]].

\bibitem{Caceres:2010rm} 
  E.~Caceres, M.~Chernicoff, A.~Guijosa and J.~F.~Pedraza,
  ``Quantum Fluctuations and the Unruh Effect in Strongly-Coupled Conformal Field Theories,''
  JHEP {\bf 1006}, 078 (2010)
  [arXiv:1003.5332 [hep-th]].

\bibitem{Hirata:2008ka} 
  T.~Hirata, S.~Mukohyama and T.~Takayanagi,
  ``Decaying D-branes and Moving Mirrors,''
  JHEP {\bf 0805}, 089 (2008)
  [arXiv:0804.1176 [hep-th]].

\bibitem{Hirayama:2010xi} 
  T.~Hirayama, P.~W.~Kao, S.~Kawamoto and F.~L.~Lin,
  ``Unruh effect and Holography,''
  Nucl.\ Phys.\ B {\bf 844}, 1 (2011)
  [arXiv:1001.1289 [hep-th]].


\bibitem{Das:2010yw} 
  S.~R.~Das, T.~Nishioka and T.~Takayanagi,
  ``Probe Branes, Time-dependent Couplings and Thermalization in AdS/CFT,''
  JHEP {\bf 1007}, 071 (2010)
  [arXiv:1005.3348 [hep-th]].
  
\bibitem{Unruh:1976db} 
  W.~G.~Unruh,
  ``Notes on black hole evaporation,''
  Phys.\ Rev.\ D {\bf 14}, 870 (1976).
  
  
  
  
  
  
  
  
  
  
  
  
  
  
  
  
  
  
  
  
  
  
  
  
  
  
  

\bibitem{Ceresole:1999zs} 
  A.~Ceresole, G.~Dall'Agata, R.~D'Auria and S.~Ferrara,
  ``Spectrum of type IIB supergravity on $AdS(5) \times T^{1,1}$: Predictions on N=1 SCFT's,''
  Phys.\ Rev.\ D {\bf 61}, 066001 (2000)
  [hep-th/9905226].
  
\bibitem{Kim:1985ez} 
  H.~J.~Kim, L.~J.~Romans and P.~van Nieuwenhuizen,
  ``The Mass Spectrum of Chiral $N=2$ $D=10$ Supergravity on $S^5$,''
  Phys.\ Rev.\ D {\bf 32}, 389 (1985).
  
  
\bibitem{Gunaydin:1984fk} 
  M.~Gunaydin and N.~Marcus,
``The Spectrum of the $S^5$ Compactification of the Chiral
$N=2$, $D=10$ Supergravity and the Unitary Supermultiplets of $U(2, 2/4)$,''
  Class.\ Quant.\ Grav.\  {\bf 2}, L11 (1985).






\bibitem{Candelas:1989js} 
  P.~Candelas and X.~C.~de la Ossa,
  ``Comments on Conifolds,''
  Nucl.\ Phys.\ B {\bf 342}, 246 (1990).

\bibitem{Minasian:1999tt} 
  R.~Minasian and D.~Tsimpis,
  ``On the geometry of nontrivially embedded branes,''
  Nucl.\ Phys.\ B {\bf 572}, 499 (2000)
  [hep-th/9911042].
  
\bibitem{Nakamura:2013yqa} 
  S.~Nakamura and H.~Ooguri,
  ``Out of Equilibrium Temperature from Holography,''
  Phys.\ Rev.\ D {\bf 88}, no. 12, 126003 (2013)
  [arXiv:1309.4089 [hep-th]].

  
\bibitem{Johnson:2003gi} 
  C.~V.~Johnson,
  ``D-branes,''
  Cambridge, USA: Univ. Pr. (2003) 548 p


  
  
\bibitem{Polchinski:1998rr} 
  J.~Polchinski,
  ``String theory. Vol. 2: Superstring theory and beyond,''
  
  
  
 
  
  
  
  


\end{thebibliography}
\end{document}